\documentclass[a4paper,11pt]{article}

\usepackage{amssymb,amsmath}

\newcommand{\ns}{\normalsize}
\newcommand{\be}{\begin{equation}}
\newcommand{\ee}{\end{equation}}
\newcommand{\ba}{\begin{eqnarray}}
\newcommand{\ea}{\end{eqnarray}}
\newcommand{\Up}{\Upsilon}
\newcommand{\la}{\lambda}
\newcommand{\si}{\sigma}
\newcommand{\ve}{\varepsilon}
\newcommand{\B}{\bar}
\newcommand{\ti}{\tilde}
\newcommand{\ph}{\phantom}
\newcommand{\nn}{\nonumber}
\newcommand{\ul}{\underline}

\numberwithin{equation}{section}

\begin{document}


\begin{titlepage}

\vspace{-3cm}

\title{\hfill{\ns hep-th/0602055\\}
   \vskip 1cm
   {\Large M-Theory on the Orbifold $\mathbb{C}^2/\mathbb{Z}_N$}\\}
   \setcounter{footnote}{0}
\author{
{\ns\large Lara B Anderson$^1$\footnote{email: anderson@maths.ox.ac.uk},
 Adam B Barrett$^2$\footnote{email: barrett@thphys.ox.ac.uk}
  \setcounter{footnote}{3}
  and Andr\'e Lukas$^2$\footnote{email: lukas@physics.ox.ac.uk}} \\[1em]
   {\ns\it$^1$Mathematical Institute, University of Oxford}\\
   {\ns 24-29 St.~Giles', Oxford OX1 3LB, UK}\\[1em]
   {\it\ns $^2$The Rudolf Peierls Centre for Theoretical Physics}\\
   {\it\ns University of Oxford}\\
   {\ns 1 Keble Road, Oxford OX1 3NP, UK}}
\date{}

\maketitle

\begin{abstract}\noindent
We construct M-theory on the orbifold $\mathbb{C}^2/\mathbb{Z}_N$ by coupling
11-dimensional supergravity to a seven-dimensional Yang-Mills theory
located on the orbifold fixed plane. It is shown that the resulting action
is supersymmetric to leading non-trivial order in the 11-dimensional
Newton constant. This action provides the starting point for a reduction
of M-theory on $G_2$ spaces with co-dimension four singularities.
\end{abstract}

\thispagestyle{empty}

\end{titlepage}


\section{Introduction}

It has been known for a long time that compactification of
11-dimensional supergravity on seven-dimensional manifolds leads to
non-chiral four-dimensional effective theories~\cite{Comp2,Comp1} and
does, therefore, not provide a viable framework for particle
phenomenology. With the construction of M-theory on the orbifold
$S^1/\mathbb{Z}_2\times\mathbb{R}^{1,9}$, Ho\v rava and
Witten~\cite{Hor-Wit} demonstrated for the first time that the
situation can be quite different for compactifications on singular
spaces.  In fact, they showed that new states in the form of two
10-dimensional $E_8$ super-Yang-Mills multiplets, located on the two
10-dimensional fixed planes of this orbifold, had to be added to the
theory for consistency and they explicitly constructed the
corresponding supergravity theory by coupling 11-dimensional
supergravity in the bulk to these super-Yang-Mills theories. It soon
became clear that this theory allows for phenomenologically
interesting Calabi-Yau compactifications~\cite{Strong}-\cite{HetM}
and, as the strong-coupling limit of the heterotic string, should be
regarded as a promising avenue towards particle phenomenology from
M-theory.

More recently, it has been discovered that phenomenologically
interesting theories can also be obtained by M-theory compactification
on singular spaces with $G_2$
holonomy~\cite{Acharya}-\cite{Acharya:2002kv}. More precisely, in such
compactifications, certain co-dimension four singularities within the
$G_2$ space lead to low-energy non-Abelian gauge
fields~\cite{Acharya,Gukov} while co-dimension seven singularities can
lead to matter fields~\cite{B,Atiyah,Gukov}. In contrast, M-theory
compactifications on $G_2$ manifolds reduce to four-dimensional ${\cal
N}=1$ theories with only Abelian vector multiplets and uncharged
chiral multiplets. Our focus in the present paper will be the
non-Abelian gauge fields arising from co-dimension four
singularities. The structure of the $G_2$ space close to such a
singularity is of the form $\mathbb{C}^2/D\times B$, where $D$ is one
of the discrete ADE subgroups of $\rm{SU}(2)$ and $B$ is a
three-dimensional space. We will, for simplicity, focus on A-type
singularities, that is $D=\mathbb{Z}_N$, which lead to gauge fields
with gauge group $\rm{SU}(N)$. A large class of singular $G_2$ spaces
containing such singularities has been obtained in
\cite{Barrett}, by orbifolding seven tori~\cite{Joyce}. It
was shown that, within this class of examples, the possible values
of $N$ are 2, 3, 4 and 6. However, in this paper, we keep $N$
general, given that there may be other constructions which lead to
more general $N$. The gauge fields are located at the fixed point
of $\mathbb{C}^2/\mathbb{Z}_N$ (the origin of $\mathbb{C}^2$) times
$B\times\mathbb{R}^{1,3}$, where $\mathbb{R}^{1,3}$ is the
four-dimensional uncompactified space-time, and are, hence,
seven-dimensional in nature. One would, therefore, expect there to
exist a supersymmetric theory which couples M-theory on the orbifold
$\mathbb{C}^2/\mathbb{Z}_N$ to seven-dimensional super-Yang-Mills
theory. It is the main purpose of the present paper to construct this
theory explicitly.

Although motivated by the prospect of applications to $G_2$
compactifications, we will formulate this problem in a slightly more
general context, seeking to understand the general structure of
low-energy M-theory on orbifolds of ADE type. Concretely, we will
construct 11-dimensional supergravity on the orbifold
$\mathbb{C}^2/\mathbb{Z}_N\times \mathbb{R}^{1,6}$ coupled to
seven-dimensional $\rm{SU}(N)$ super-Yang-Mills theory located on the
orbifold fixed plane $\{\bf{0}\}\times\mathbb{R}^{1,6}$. For ease of
terminology, we will also refer to this orbifold plane, somewhat
loosely as the ``brane''. This result can then be applied to
compacitifications of M-theory on $G_2$ spaces with
$\mathbb{C}^2/\mathbb{Z}_N$ singularities, as well as to other
problems (for example M-theory on certain singular limits of K3). We
stress that this construction is very much in the spirit of the Ho\v
rava-Witten theory~\cite{Hor-Wit}, which couples 11-dimensional
supergravity on $S^1/\mathbb{Z}_2\times \mathbb{R}^{1,9}$ to
10-dimensional super-Yang-Mills theory.

Let us briefly outline our method to construct this theory which
relies on combining information from the known actions of
11-dimensional~\cite{Julia,GSW} and seven-dimensional
supergravity~\cite{Berghshoeff}-\cite{Park}. Firstly, we constrain the field content of
11-dimensional supergravity (the ``bulk fields'') to be compatible
with the $\mathbb{Z}_N$ orbifolding symmetry. We will call the
Lagrangian for this constrained version of 11-dimensional supergravity
${\cal L}_{11}$. As a second step this action is truncated to seven
dimensions, by requiring all fields to be independent of the
coordinates $y$ of the orbifold $\mathbb{C}^2/\mathbb{Z}_N$ (or,
equivalently, constraining it to the orbifold plane at $y=0$). The
resulting Lagrangian, which we call ${\cal L}_{11}|_{y=0}$, is
invariant under half of the original 32 supersymmetries and represents
a seven-dimensional ${\cal N}=1$ supergravity theory which turns out
to be coupled to a single $U(1)$ vector multiplet for $N>2$ or three
$U(1)$ vector multiplets for $N=2$. As a useful by-product, we obtain
an explicit identification of the (truncated) 11-dimensional bulk
fields with the standard fields of 7-dimensional Einstein-Yang-Mills
(EYM) supergravity. We know that the additional states on the
orbifold fixed plane should form a seven-dimensional vector multiplet
with gauge group ${\rm SU}(N)$.  In a third step, we couple these
additional states to the truncated seven-dimensional bulk theory
${\cal L}_{11}|_{y=0}$ to obtain a seven-dimensional EYM supergravity
${\cal L}_{SU(N)}$ with gauge group $U(1)\times SU(N)$ for $N>2$ or
$U(1)^3\times SU(N)$ for $N=2$. We note that, given a fixed gauge group
the structure of ${\cal L}_{SU(N)}$ is essentially determined by
seven-dimensional supergravity. We further write this theory in a form
which explicitly separates the bulk degrees of freedom (which we have
identified with 11-dimensional fields) from the degrees of freedom in
the $SU(N)$ vector multiplets. Given this preparation we prove in
general that the action
\begin{equation}
S_{11-7}=\int_{\mathcal{M}_{11}} \mathrm{d}^{11}x \left[ \mathcal{L}_{11} + \delta^{(4)}
(y^A)\left( \mathcal{L}_{\mathrm{SU(N)}}-\kappa^{8/9}\mathcal{L}_{11}\right) \right]
\label{S117}
\end{equation}
is supersymmetric to leading non-trivial order in an expansion in
$\kappa$, the 11-dimensional Newton constant. Inserting the various
Lagrangians with the appropriate field identifications into this
expression then provides us with the final result.

The plan of the paper is as follows. In Section \ref{Bulk} we remind
the reader of the action of 11-dimensional supergravity. As
mentioned above this is to be our bulk theory. We then go on to
discuss the constraints that arise on the fields from putting this
theory on the orbifold. We also lay out our conventions for rewriting
11-dimensional fields according to a seven plus four split of the
coordinates. In Section \ref{reduction} we examine our bulk Lagrangian
constrained to the orbifold plane and recast it in standard
seven-dimensional form. The proof that the action~\eqref{S117} is
indeed supersymmetric to leading non-trivial order is presented in
Section \ref{construct}. Finally, in Section \ref{results} we present
the explicit result for the coupled 11-/7-dimensional action and the
associated supersymmetry transformations. We end with a discussion of
our results and an outlook on future directions. Three appendices
present technical background material. In Appendix \ref{spinors} we
detail our conventions for spinors in eleven, seven and four
dimensions and describe how we decompose 11-dimensional
spinors. We also give some useful spinor identities. In Appendix
\ref{Pauli} we have collected some useful group-theoretical
information related to the cosets $SO(3,M)/SO(3)\times SO(M)$ of $d=7$
EYM supergravity which will be used in the reduction of the bulk
theory to seven-dimensions. The final Appendix is a self-contained
introduction to EYM supergravity in seven dimensions.


\section{Eleven-dimensional supergravity on the orbifold} \label{Bulk}
In this section we begin our discussion of M-theory on
$\mathcal{M}^N_{11}=\mathbb{R}^{1,6}\times\mathbb{C}^2/\mathbb{Z}_N$
by describing the bulk action and the associated bulk supersymmetry
transformations. We recall that fields propagating on orbifolds are
subject to certain constraints on their configurations and proceed by
listing and explaining these. First however we lay out our
conventions, and briefly describe the decomposition of spinors
in a four plus seven split of the coordinates.\\

We take space-time to have mostly positive signature, that is
$(-++\ldots+)$, and use indices $M,N,\ldots=0,1,\ldots,10$ to label
the 11-dimensional coordinates $(x^M)$. It is often convenient to
split these into four coordinates $y^A$, where $A,B,\ldots=7,8,9,10$,
in the directions of the orbifold $\mathbb{C}^2/\mathbb{Z}_N$ and seven
remaining coordinates $x^\mu$, where $\mu,\nu,\ldots=0,1,2,\ldots,6$,
on $\mathbb{R}^{1,6}$.  Frequently, we will also use complex
coordinates $(z^p,\bar{z}^{\bar{p}})$ on $\mathbb{C}^2/\mathbb{Z}_N$,
where $p,q,\ldots=1,2$, and $\bar{p},\bar{q},\ldots=\bar{1},\B{2}$
label holomorphic and anti-holomorphic coordinates, respectively.
Underlined versions of all the above index types denote the associated
tangent space indices.

All 11-dimensional spinors in this paper are Majorana. Having
split coordinates into four- and seven-dimensional parts it is useful
to decompose 11-dimensional Majorana spinors accordingly as tensor
products of $SO(1,6)$ and $SO(4)$ spinors. To this end, we introduce a
basis of left-handed $SO(4)$ spinors $\{\rho^i\}$ and their
right-handed counterparts $\{\rho^{\bar{\jmath}}\}$ with indices
$i,j,\ldots =1,2$ and $\bar{\imath},\bar{\jmath},\ldots
=\bar{1},\bar{2}$. Up to an overall rescaling, this basis can be
defined by the relations $\gamma^{\underline{A}}\rho^i =
\left(\gamma^{\underline{A}}\right)_{\B{\jmath}}^{\ph{j}i}\rho^{\B{\jmath}}$.
An 11-dimensional spinor $\psi$ can then be written as
\begin{equation}
\psi=\psi_i (x,y)\otimes\rho^i + \psi_{\bar{\jmath}}(x,y)\otimes\rho^{\bar{\jmath}},
 \label{spinor7}
\end{equation}
where the 11-dimensional Majorana condition on $\psi$ forces
$\psi_i(x,y)$ and $\psi_{\bar{\jmath}}(x,y)$ to be $SO(1,6)$
symplectic Majorana spinors. In the following, for any 11-dimensional
Majorana spinor we will denote its associated seven-dimensional 
symplectic Majorana spinors by the same symbol with additional $i$
and $\bar{\imath}$ indices. A full account of spinor conventions used
in this paper, together with a derivation of the above decomposition
can be found in Appendix \ref{spinors}.\\

With our conventions in place, we proceed by reviewing 11-dimensional
supergravity~\cite{Julia,GSW,Hor-Wit}.  Its field content consists of the vielbein
${e_M}^{\underline{M}}$ and associated metric
$g_{MN}=\eta_{\underline{M}\underline{N}}e_M^{\phantom{M}\underline{M}}
e_N^{\phantom{N}\underline{N}}$,
the three-form field $C$, with field strength $G=\mathrm{d}C$, and the
gravitino $\Psi_M$. In this paper we shall not compute any four fermi
terms (and associated cubic fermion terms in the supersymmetry
transformations), and so we shall neglect them throughout. The action
is given by
\begin{eqnarray} \label{11dsugra}
\mathcal{S}_{11} & = &  \frac{1}{\kappa^2}\int_{\mathcal{M}^N_{11}}\mathrm{d}^{11}x\sqrt{-g}
 \bigg( \frac{1}{2}R-\frac{1}{2}\bar{\Psi}_M\Gamma^{MNP}\nabla_N\Psi_P-\frac{1}{96}G_{MNPQ}G^{MNPQ}
 \nonumber \\ 
 & & \hspace{3.5cm} -\frac{1}{192}\Big(\bar{\Psi}_M\Gamma^{MNPQRS}\Psi_S+12\bar{\Psi}^N
     \Gamma^{PQ}\Psi^R\Big)
 G_{NPQR}\bigg) \\
& & -\frac{1}{12\kappa^2}\int_{\mathcal{M}^N_{11}}C\wedge G\wedge G +\dots\nonumber
\end{eqnarray}
where the dots stand for terms quartic in the gravitino.  Here $\kappa$
is the 11-dimensional Newton constant, $\Gamma^{M_1M_2\ldots M_n}$
denote anti-symmetrized products of gamma matrices in the usual way,
and $\nabla_I$ is the spinor covariant derivative, defined in terms of
the spin connection $\omega$ by
\begin{equation}
\nabla_M=\partial_M+\frac{1}{4}\omega_M^{\phantom{M}\underline{M}\underline{N}}
\Gamma_{\underline{M}\underline{N}}.
\end{equation}
The transformation laws of local supersymmetry, parameterized by the spinor $\eta$, read
\begin{eqnarray}
\delta e_M^{\phantom{M}\underline{N}} & = & \bar{\eta}\Gamma^{\underline{N}}\Psi_{M} \nn \\
\delta C_{MNP} & = & -3\bar{\eta}\Gamma_{[MN}\Psi_{P]} \label{11dsusy1} \\
\delta \Psi_M & = & 2\nabla_M\eta+\frac{1}{144}\left( \Gamma_M^{\phantom{M}NPQR}
 -8\delta_M^N\Gamma^{PQR}\right)\eta G_{NPQR} + \dots ,  \nn
\end{eqnarray}
where the dots denote terms cubic in the gravitino.\\

In order for the above bulk theory to be consistent on the orbifold
$\mathbb{C}^2/\mathbb{Z}_N\times\mathbb{R}^{1,6}$ we need to constrain
fields in accordance with the $\mathbb{Z}_N$ orbifold action. Let us
now discuss in detail how this works.  

We denote by $R$ the $SO(4)$ matrix of order $N$ that generates the
$\mathbb{Z}_N$ symmetry on our orbifold. This matrix acts on the
11-dimensional coordinates as $(x,y)\rightarrow (x,Ry)$ which
implies the existence of a seven-dimensional fixed plane characterized
by $\{ y=0\}$. For a field $X$ to be well-defined on the orbifold
it must satisfy
\begin{equation} \label{theta}
X(x,Ry)=\Theta(R)X(x,y)
\end{equation}
for some linear operator $\Theta(R)$ that represents the generator of
$\mathbb{Z}_N$. In constructing our theory we have to choose, for each
field, a representation $\Theta$ of $\mathbb{Z}_N$ for which we wish
to impose this constraint. For the theory to be well-defined, these
choices of representations must be such that the action
\eqref{11dsugra} is invariant under the $\mathbb{Z}_N$ orbifold
symmetry. Concretely, what we do is choose how each index type
transforms under $\mathbb{Z}_N$. We take $R\equiv (R^A_{\ph{A}B})$ to
be the transformation matrix acting on curved four-dimensional indices
$A,B,\ldots$ while the generator acting on tangent space indices
$\underline{A},\underline{B},\ldots$ is some other $SO(4)$ matrix
$T^{\underline{A}}_{\phantom{A}\underline{B}}$. It turns out that this
matrix must be of order $N$ for the four-dimensional components of the
vielbein to remain non-singular at the orbifold fixed
plane. Seven-dimensional indices $\mu ,\nu ,\ldots$ transform
trivially. Following the correspondence Eq.~\eqref{spinor7},
11-dimensional Majorana spinors $\psi$ are described by two pairs
$\psi_i$ and $\psi_{\bar{\imath}}$ of seven-dimensional symplectic
Majorana spinor.  We should, therefore, specify how the $\mathbb{Z}_N$
symmetry acts on indices of type $i$ and $\bar{\imath}$. Supersymmetry
requires that at least some spinorial degrees of freedom survive at
the orbifold fixed plane. For this to be the case, one of these type
of indices, $i$ say, must transform trivially. Invariance of fermionic
terms in the action \eqref{11dsugra} requires that the other indices,
that is those of type $\bar{\imath}$, be acted upon by a $U(2)$
matrix $S_{\bar{\imath}}^{\phantom{i}\bar{\jmath}}$ that satisfies the equation
\begin{equation} \label{STconstraint}
{S_{\bar{\imath}}}^{\bar{k}}\left(\gamma^{\underline{A}}\right)_{\B{k}}^{\ph{k}\jmath}=
{T^{\ul{A}}}_{\underline{B}}
{\left(\gamma^{\underline{B}}\right)_{\B{\imath}}}^{\jmath}.
\end{equation}
Given this basic structure, the constraints satisfied by the fields are as follows
\begin{eqnarray} 
e_\mu^{\ph{\mu}\underline{\nu}}(x,Ry)&=&
e_\mu^{\ph{\mu}\underline{\nu}}(x,y), \label{cond1} \\
e_A^{\ph{A}\underline{\nu}}(x,Ry)&=&
(R^{-1})_A^{\phantom{A}B}e_B^{\ph{B}\underline{\nu}}(x,y),\\
e_\mu^{\ph{\mu}\underline{A}}(x,Ry)&=&T^{\underline{A}}_{\phantom{A}\underline{B}}
e_\mu^{\ph{\mu}\underline{B}}(x,y),\\
e_A^{\ph{A}\underline{B}}(x,Ry)&=&(R^{-1})_A^{\phantom{A}C}T^{\underline{B}}_{\phantom{B}\underline{D}}
e_C^{\ph{C}\underline{D}}(x,y), \label{cond4}
\end{eqnarray}
\begin{eqnarray}
C_{\mu\nu\rho}(x,Ry) & = & C_{\mu\nu\rho}(x,y), \\ \label{Ccond} C_{\mu\nu A}(x,Ry) & = &
(R^{-1})_A^{\phantom{A}B}C_{\mu\nu B}(x,y), \: \:\: \mathrm{etc.} \label{cond6}\\
\Psi_{\mu i}(x,Ry)&=&\Psi_{\mu i}(x,y), \\ \Psi_{\mu
\bar{\imath}}(x,Ry)&=&S_{\bar{\imath}}^{\phantom{i}\bar{\jmath}}\Psi_{\mu
\bar{\jmath}}(x,y),\\ \Psi_{A
i}(x,Ry)&=&(R^{-1})_{A}^{\phantom{A}B}\Psi_{B i}(x,y),\\ \Psi_{A
\bar{\imath}}(x,Ry)&=&(R^{-1})_{A}^{\phantom{A}B}S_{\bar{\imath}}^{\phantom{i}\bar{\jmath}}\Psi_{B
\bar{\jmath}}(x,y). \label{condn} \ea
Furthermore, covariance of the supersymmetry transformation laws with
respect to $\mathbb{Z}_N$ requires
\ba \eta_{i}(x,Ry)&=&\eta_{i}(x,y), \\
\eta_{\bar{\imath}}(x,Ry)&=&S_{\bar{\imath}}^{\phantom{i}\bar{\jmath}}\eta_{\bar{\jmath}}(x,y).
\ea
In complex coordinates $(z^p,\bar{z}^{\bar{p}})$, it is convenient to represent $R$
by the following matrices
\begin{equation} \label{R}
(R^p_{\ph{p}q})=e^{2i\pi /N}\boldsymbol{1}_2, \hspace{0.3cm} (R^{\B{p}}_{\ph{p}\B{q}})=
e^{-2i\pi /N}\boldsymbol{1}_2, \hspace{0.3cm} (R^{\B{p}}_{\ph{p}q})=(R^p_{\ph{p}\B{q}})=0\, . 
\end{equation}
Using this representation, the constraint~\eqref{cond4} implies
\begin{equation}
 {e_p}^{\ul{A}}=e^{-2i\pi /N}{T^{\ul{A}}}_{\ul{B}}\, {e_p}^{\ul{B}}\; .
\end{equation}
Hence, for the vierbein ${e_{\ul{A}}}^{\ul{B}}$ to be non-singular $T$
must have two eigenvalues $e^{2i\pi /N}$. Similarly, the conjugate of the
above equation shows that $T$ should have two eigenvalues $e^{-2i\pi /N}$.
Therefore, in an appropriate basis we can use the following representation
\begin{equation}
 ({T^{\ul{p}}}_{\ul{q}})=e^{2i\pi /N}{\bf 1}_2\; ,\qquad
 ({T^{\ul{\bar p}}}_{\ul{\bar q}})=e^{-2i\pi /N}{\bf 1}_2\; .\label{T}
\end{equation}
Given these representations for $R$ and $T$, the matrix $S$ is uniquely fixed
by Eq.~\eqref{STconstraint} to be
\begin{equation}
 ({S_{\ul{i}}}^{\ul{j}})=e^{2i\pi /N}{\bf 1}_2\; .\label{S}
\end{equation}
We will use the explicit form of $R$, $T$ and $S$ above to analyze the
degrees of freedom when we truncate fields to be $y$ independent.

When the 11-dimensional fields are taken to be independent of the
orbifold $y$ coordinates, the constraints~\eqref{cond1}--\eqref{condn}
turn into projection conditions, which force certain field components
to vanish. As we will see shortly, the surviving field components fit
into seven-dimensional ${\cal N}=1$ supermultiplets, a confirmation
that we have chosen the orbifold $\mathbb{Z}_N$ action on fields
compatible with supersymmetry. More precisely, for the case $N>2$, we
will find a seven-dimensional gravity multiplet and a single
$U(1)$ vector multiplet. Hence, we expect the associated
seven-dimensional ${\cal N}=1$ Einstein-Yang-Mills (EYM) supergravity
to have gauge group $U(1)$.  For $\mathbb{Z}_2$ the situation is
slightly more complicated, since, unlike for $N>2$, some of the field
components which transform bi-linearly under the generators are now
invariant.  This leads to two additional vector multiplets, so that
the associated theory is a seven-dimensional ${\cal N}=1$ EYM
supergravity with gauge group $U(1)^3$.  In the following section, we
will write down this seven-dimensional theory, both for $N=2$ and
$N>2$, and find the explicit identifications of truncated
11-dimensional fields with standard seven-dimensional supergravity
fields.


\section{Truncating the bulk theory to seven dimensions} \label{reduction}
In this section, we describe in detail how the bulk theory is
truncated to seven dimensions. We recall from the introduction that
this constitutes one of the essential steps in the construction of the
theory. As a preparation, we explicitly write down
the components of the 11-dimensional fields that survive on the
orbifold plane and work out how these fit into seven-dimensional
super-multiplets. We then describe, for each orbifold, the
seven-dimensional EYM supergravity with the appropriate field
content. By an explicit reduction of the 11-dimensional theory and
comparison with this seven-dimensional theory, we find a list of
identification between 11- and 7-dimensional fields which is
essential for our subsequent construction.\\

To discuss the truncated field content, we use the
representations~\eqref{R}, \eqref{T}, \eqref{S} of $R$, $T$ and $S$
and the orbifold conditions \eqref{cond1}-\eqref{condn} for $y$
independent fields. For $N>2$ we find that the
surviving components are given by $g_{\mu\nu}$,
$e_p^{\ph{p}\underline{q}}$, $C_{\mu\nu\rho}$, $C_{\mu p\B{q}}$,
$\Psi_{\mu i}$, $(\Gamma^p\Psi_p)_i$ and
$(\Gamma^{\B{p}}\Psi_{\B{p}})_i$. Meanwhile, the spinor $\eta$ which
parameterizes supersymmetry reduces to $\eta_i$, a single symplectic
Majorana spinor, which corresponds to seven-dimensional ${\cal N}=1$
supersymmetry. Comparing with the structure of seven-dimensional
multiplets (see Appendix~\ref{bigEYM} for a review of
seven-dimensional EYM supergravity), these field components fill out
the seven-dimensional supergravity multiplet and a single $U(1)$
vector multiplet.  For the case of the $\mathbb{Z}_2$ orbifold, a
greater field content survives, corresponding in seven-dimensions to a
gravity multiplet plus three $U(1)$ vector multiplets. The surviving
fields in this case are expressed most succinctly by $g_{\mu\nu}$,
$e_A^{\ph{A}\underline{B}}$, $C_{\mu\nu\rho}$, $C_{\mu AB}$,
$\Psi_{\mu i}$ and $\Psi_{A\B{\imath}}$. The spinor $\eta$ which
parameterizes supersymmetry again reduces to $\eta_i$, a single
symplectic Majorana spinor.

These results imply that the truncated bulk theory is a
seven-dimensional $\mathcal{N}=1$ EYM supergravity with gauge group
$U(1)^n$, where $n=1$ for $N>2$ and $n=3$ for $N=2$. In the following,
we discuss both cases and, wherever possible, use a unified
description in terms of $n$, which can be set to either $1$ or $3$, as
appropriate.  The correspondence between 11-dimensional truncated
fields and seven-dimensional supermultiplets is as follows. The
gravity super-multiplet contains the purely seven-dimensional parts of
the 11-dimensional metric, gravitino and three-form, that is,
$g_{\mu\nu}$, $\Psi_{\mu i}$ and $C_{\mu\nu\rho}$, along with three
vectors from $C_{\mu AB}$, a spinor from $\Psi_{A\B{\imath}}$ and the
scalar $\mathrm{det}(e_A^{\ph{A}\underline{B}})$. The remaining
degrees of freedom, that is, the remaining vector(s) from $C_{\mu
AB}$, the remaining spinor(s) from $\Psi_{A\B{\imath}}$ and the
scalars contained in
$v_A^{\ph{A}\underline{B}}:=\mathrm{det}(e_A^{\ph{A}\underline{B}})^{-1/4}e_A^{\ph{A}\underline{B}}$,
the unit-determinant part of $e_A^{\ph{A}\underline{B}}$, fill out $n$
seven-dimensional vector multiplets. The $n+3$ Abelian gauge fields
transform under the $SO(3,n)$ global symmetry of the $d=7$ EYM
supergravity while the vector multiplet scalars parameterize the coset
$SO(3,n)/SO(3)\times SO(n)$. Let us describe how such coset spaces are
obtained from the vierbein $v_A^{\ph{A}\underline{B}}$, starting with
the generic case $N>2$ with seven-dimensional gauge group $U(1)$, that
is, $n=1$. In this case, the rescaled vierbein
$v_A^{\ph{A}\underline{B}}$ reduces to $v_p^{\ph{p}\underline{q}}$,
which represents a set of $2\times 2$ matrices with determinant one,
identified by $SU(2)$ transformations acting on the tangent space
index. Hence, these matrices form the coset $SL(2,\mathbb{C})/SU(2)$
which is isomorphic to $SO(3,1)/SO(3)$, the correct coset
space for $n=1$. For the special $\mathbb{Z}_2$ case, which implies
$n=3$, the whole of $v_A^{\ph{A}\underline{B}}$ is present and forms
the coset space $SL(4,\mathbb{R})/SO(4)$. This space is
isomorphic to $SO(3,3)/SO(3)^2$ which is indeed the correct
coset space for $n=3$.

We now briefly review seven-dimensional EYM supergravity with
gauge group $U(1)^n$. A more general account of seven-dimensional
supergravity including non-Abelian gauge groups can be found in
Appendix \ref{bigEYM}. The seven-dimensional ${\cal N}=1$ supergravity multiplet
contains the vielbein $\ti{e}_{\mu}^{\ph{\mu}\underline{\nu}}$, the
gravitino $\psi_{\mu i}$, a triplet of vectors $A_{\mu i}^{\ph{\mu
i}j}$ with field strengths $F_i^{\ph{i}j}=\mathrm{d}A_i^{\ph{i}j}$, a
three-form $\ti{C}$ with field strength $\ti{G}=\mathrm{d}\ti{C}$, a
spinor $\chi_i$, and a scalar $\si$. A seven-dimensional
vector multiplet contains a $U(1)$ gauge field $A_\mu$ with field
strength $F=\mathrm{d}A$, a gaugino $\la_i$ and a triplet of scalars
$\phi_i^{\ph{i}j}$. Here, all spinors are symplectic Majorana spinors
and indices $i,j,\ldots=1,2$ transform under the $SU(2)$
R-symmetry. For ease of notation, the three vector fields in the
supergravity multiplet and the $n$ additional Abelian gauge fields from
the vector multiplet are combined into a single $SO(3,n)$ vector
$A_\mu^I$, where $I,J,\ldots =1,\ldots ,n+3$. The coset space
$SO(3,n)/SO(3)\times SO(n)$ is described by a $(3+n)\times (3+n)$ matrix
$\ell_{I}^{\ph{I}\underline{J}}$, which depends on the $3n$ vector
multiplet scalars and satisfies the $SO(3,n)$ orthogonality condition
\begin{equation}
\ell_{I}^{\ph{I}\underline{J}}\ell_{K}^{\ph{K}\underline{L}}\eta_{\underline{J}\underline{L}}=\eta_{IK}
\end{equation}
with
$(\eta_{IJ})=(\eta_{\underline{I}\underline{J}})=\rm{diag}(-1,-1,-1,+1,\ldots,+1)$. Here,
indices $I,J,\ldots=1,\ldots,(n+3)$ transform under $SO(3,n)$. Their
flat counterparts $\underline{I},\underline{J}\ldots$ decompose into a
triplet of $SU(2)$, corresponding to the gravitational directions and
$n$ remaining directions corresponding to the vector multiplets. Thus
we can write $\ell_{I}^{\ph{I}\underline{J}}\to
(\ell_{I}^{\ph{I}u},\ell_I^{\ph{I}\alpha})$, where $u=1,2,3$ and
$\alpha=1,\ldots,n$. The adjoint $SU(2)$ index $u$ can be converted
into a pair of fundamental $SU(2)$ indices by multiplication with the
Pauli matrices, that is,
\begin{equation}
\ell_{I\ph{i}j}^{\ph{I}i}=\frac{1}{\sqrt{2}}\ell_I^{\ph{I}u}(\si_u)^i_{\ph{i}j}.
\end{equation}
The Maurer-Cartan forms $p$ and $q$ of the matrix $\ell$, defined by
\ba 
p_{\mu\alpha \ph{i}j}^{\ph{\mu\alpha }i}&=&\ell^{I}_{\ph{I}\alpha}
\partial_\mu \ell_{I\ph{i}j}^{\ph{\mu}i}, \label{Maurer1}\\
q_{\mu \ph{i}j\ph{k}l}^{\ph{\mu}i\ph{j}k}&=&\ell^{Ii}_{\ph{Ii}j}
 \partial_\mu \ell_{I\ph{k}l}^{\ph{\mu}k}, \\
q_{\mu \phantom{i}j}^{\phantom{\mu}i}&=&\ell^{Ii}_{\ph{Ii}k}
\partial_\mu \ell_{I\ph{k}j}^{\ph{\mu}k}, \label{Maurern}
\ea
will be needed as well.

With everything in place, we can now write down our expression for $\mathcal{L}_7^{(n)}$,
the Lagrangian of seven-dimensional ${\cal N}=1$ EYM supergravity with gauge group
$U(1)^n$ \cite{Park}. Neglecting four-fermi terms, it is given by
\begin{eqnarray} \label{7dsugra}
\mathcal{L}_7^{(n)}\!\!\! &=&\!\!\!\frac{1}{\kappa^2_7}\sqrt{-\ti{g}}\left\{\frac{1}{2}
R-\frac{1}{2}\bar{\psi}_{\mu
}^{i}\Upsilon ^{\mu \nu \rho }\hat{\mathcal{D}} _{\nu }\psi _{\rho i}-\frac{1}{4}e^{-2\si}
\left( \ell_{I\phantom{i}j}^{\phantom{I}i}\ell_{%
J\phantom{j}i}^{\phantom{J}j}+\ell_{I}^{\ph{I}\alpha}\ell_{J\alpha}\right)
F_{\mu \nu }^{I}F^{J\mu \nu } \right.  \notag \\
&&\hspace{1.5cm}-\frac{1}{96}e^{4\si}\ti{G}_{\mu \nu \rho \sigma }\ti{G}^{\mu \nu \rho \sigma }
-\frac{1}{2}\bar{\chi}^{i}\Upsilon ^{\mu }\hat{\mathcal{D}} _{\mu }\chi _{i}-\frac{5}{%
2}\partial _{\mu }\si \partial ^{\mu }\si +\frac{\sqrt{5}}{2}\left( \bar{\chi}^{i}
\Upsilon ^{\mu \nu }\psi _{\mu i}+\bar{\chi}^{i}\psi _{i}^{\nu }\right)
\partial _{\nu }\si   \notag \\
&&\hspace{1.5cm}-\frac{1}{2}\bar{\lambda}^{\alpha i}\Upsilon ^{\mu }\hat{\mathcal{D}}
_{\mu }\lambda _{\alpha i}-\frac{1}{2}p_{\mu\alpha \phantom{i}j}^{\phantom{\mu\alpha}i}p_{%
\phantom{\mu\alpha j}i}^{\mu\alpha j}-\frac{1}{\sqrt{2}}\left( \bar{\lambda}^{\alpha i}
\Upsilon ^{\mu \nu }\psi _{\mu
j}+\bar{\lambda}^{\alpha i}\psi _{j}^{\nu }\right) p_{\nu\alpha \phantom{j}i}^{%
\phantom{\nu\alpha}j}  \notag \\
&&\hspace{1.5cm}+e^{2\si}\ti{G}_{\mu \nu \rho \sigma }\left[ 
\frac{1}{192}\left( 12\bar{\psi}^{\mu i}\Upsilon ^{\nu \rho }\psi _{i}^{\si}+ \bar{\psi}_{\la}^{i}
\Upsilon ^{\lambda \mu \nu \rho \sigma \tau }\psi _{\tau
i}\right)+\frac{1}{48\sqrt{5}}\left( 4\bar{\chi}^{i}\Upsilon
^{\mu \nu \rho }\psi _{i}^{\si} \right.\right.\nn 
\end{eqnarray}
\begin{eqnarray}
&& \hspace{3.5cm} \left. \left. -\bar{\chi}^{i}\Upsilon ^{\mu \nu \rho \sigma
\tau }\psi _{\tau i}\right) -\frac{1}{320}\bar{\chi}^{i}\Upsilon ^{\mu \nu
\rho \sigma }\chi _{i}+\frac{1}{192}\bar{\lambda}^{\alpha i}\Upsilon ^{\mu \nu \rho\sigma }
\lambda _{\alpha i}\right] \nn \\
&&\hspace{1.5cm} -ie^{-\si}F_{\mu \nu }^{I}\ell_{I%
\phantom{j}i}^{\phantom{I}j}\left[ \frac{1}{4\sqrt{2}}\left( \bar{\psi}%
_{\rho }^{i}\Upsilon ^{\mu \nu \rho \sigma }\psi _{\sigma j}+2\bar{\psi}%
^{\mu i}\psi _{j}^{\nu }\right) +\frac{1}{2\sqrt{10}}\left( \bar{\chi}%
^{i}\Upsilon ^{\mu \nu \rho }\psi _{\rho j}-2\bar{\chi}^{i}\Upsilon ^{\mu
}\psi _{j}^{\nu }\right) \right.  \notag \\
&&\hspace{3.9cm}\left. +\frac{3}{20\sqrt{2}}\bar{\chi}^{i}\Upsilon ^{\mu \nu
}\chi _{j}-\frac{1}{4\sqrt{2}}\bar{\lambda}^{\alpha i}\Upsilon ^{\mu \nu }\lambda
_{\alpha j}\right]  \notag \\
&&\hspace{1.5cm}+e^{-\si }F_{\mu \nu }^{I}\ell_{I\alpha}\left[ 
\frac{1}{4}\left( 2\bar{\lambda}^{\alpha i}\Upsilon ^{\mu }\psi _{i}^{\nu }
-\bar{\lambda}^{\alpha i}\Upsilon ^{\mu \nu \rho }\psi _{\rho i}\right)
 +\frac{1}{2\sqrt{5}}\bar{\lambda}^{\alpha i}\Upsilon ^{\mu \nu }\chi _{i}\right]  \notag \\
&&\hspace{1.5cm}\left. -\frac{1}{96}\epsilon^{\mu\nu\rho\sigma\kappa\lambda\tau}C_{\mu\nu\rho
}F_{\sigma\kappa}^{\tilde{I}}F_{\tilde{I}}{}_{\lambda\tau} \right\}\; .
\end{eqnarray}
In this Lagrangian the covariant derivatives of symplectic Majorana spinors $\epsilon _{i}$
are defined by 
\begin{equation}
\hat{\mathcal{D}} _{\mu }\epsilon _{i}=\partial _{\mu }\epsilon _{i}+\frac{1}{2}q_{\mu
i}^{\phantom{\mu i}j}\epsilon _{j}+\frac{1}{4}\ti{\omega} _{\mu }^{\phantom{\mu}%
\underline{\mu }\underline{\nu }}\Upsilon _{\underline{\mu }\underline{\nu }%
}\epsilon _{i}.
\end{equation}
The associated supersymmetry transformations, parameterized by the spinor $\varepsilon_i$,
are, up to cubic fermion terms, given by
\ba
\delta \sigma &=& \frac{1}{\sqrt{5}}\bar{\chi}^{i}\varepsilon _{i}\; ,  \nn\\
\delta \ti{e} _{\mu }^{%
\underline{\nu}}&=&\B{\varepsilon }^{i}\Upsilon ^{\underline{\nu}}\psi
_{\mu i }\; , \nn\\
\delta \psi _{\mu i}&=&2\hat{\mathcal{D}} _{\mu }\varepsilon _{i}-\frac{e^{2\sigma}}{80}
\left(\Upsilon _{\mu }^{\ph{\mu} \nu \rho \sigma \eta }-\frac{8}{3}\delta _{\mu }^{\nu
}\Upsilon ^{\rho \sigma \eta }\right) \varepsilon _{i}\ti{G}_{\nu \rho \sigma \eta
}+\frac{ie^{-\sigma}}{5\sqrt{2}}\left(
\Upsilon _{\mu }^{\ph{\mu}\nu \rho }-8\delta _{\mu }^{\nu }\Upsilon ^{\rho
}\right) \varepsilon _{j}F_{\nu \rho }^{I}\ell_{I\phantom{i}i}^{\phantom{I}j}\; , \nn\\
\delta \chi _{i}&=& \sqrt{5}\Upsilon ^{\mu }\varepsilon _{i}\partial _{\mu }\si -
\frac{1}{24\sqrt{5}}\Upsilon ^{\mu \upsilon \rho \sigma }\varepsilon
_{i}\ti{G}_{\mu \nu \rho \sigma }e^{2\si}\text{ }-\frac{i}{\sqrt{10}}\Upsilon ^{\mu\nu }
\varepsilon _{j}F_{\mu\nu }^{I}\ell_{I\phantom{i}i}^{\phantom{I}j}e^{-\si}, \nn\\
\delta \ti{C}_{\mu \nu \rho }&=&\left( -3\B{\psi }^{i}_{\left[ \mu \right. }\Upsilon _{\left.
 \nu \rho \right] }\varepsilon _{i}-\frac{2}{\sqrt{5}}\B{\chi }^{i}\Upsilon _{\mu \nu \rho }
\varepsilon_{i}\right) e^{-2\si},  \\
\ell_{I\phantom{i}j}^{\phantom{I}i}\delta A_{\mu }^{I}&=&\left[ i\sqrt{2}\left( 
\B{\psi }_{\mu }^{i}\varepsilon _{j}-\frac{1}{2}%
\delta _{j}^{i}\B{\psi }_{\mu }^{k}\varepsilon _{k}\right) -\frac{2i}{%
\sqrt{10}}\left( \B{\chi }^{i}\Upsilon _{\mu }\varepsilon _{j}-\frac{1%
}{2}{}\delta _{j}^{i}\B{\chi }^{k}\Upsilon _{\mu }\varepsilon
_{k}\right) \right] e^{\si},  \nn\\
\ell_{I}^{\ph{I}\alpha}\delta A_{\mu }^{I}&=&\B{\varepsilon }%
^{i}\Upsilon _{\mu }\lambda _{i}^\alpha e^{\si},  \nn\\
\delta \ell_{I\phantom{i}j}^{\phantom{I}i}&=&-i\sqrt{2}\B{\varepsilon }^{i}
\lambda _{\alpha j}\ell_{I}^{\ph{I}\alpha}+\frac{i}{\sqrt{2}}\B{\varepsilon }^{k}
\lambda _{\alpha k}\ell_{I}^{\ph{I}\alpha}\delta _{j}^{i}\; ,  \nn\\
\delta \ell_{I}^{\ph{I}\alpha}&=&-i\sqrt{2}\B{\varepsilon }^{i}\lambda_{j}^\alpha
\ell_{I\phantom{j}i}^{\phantom{I}j}\; , \nn\\
\delta \lambda _{i}^\alpha &=&-\frac{1}{2}\Upsilon ^{\mu\nu
}\varepsilon _{i}F_{\mu\nu}^{I}\ell_{I}^{\ph{I}\alpha}e^{-\si}+\sqrt{2}i
\Upsilon ^{\mu }\varepsilon _{j}p_{\mu \phantom{\alpha i}j}^{\phantom{\mu}\alpha i}\; . \nn
\ea
\\
We now explain in detail how the truncated  bulk theory corresponds
to the above seven-dimensional EYM supergravity with gauge group $U(1)^n$,
where $n=1$ for the $\mathbb{Z}_N$ orbifold with $N>2$ and $n=3$ for the
special $\mathbb{Z}_2$ case. It is convenient to choose the seven-dimensional
Newton constant $\kappa_7$ as $\kappa_7=\kappa^{5/9}$. The correspondence
between 11- and 7-dimensional Lagrangians can then be written as
\begin{equation} \label{equivalence}
\kappa ^{8/9}\mathcal{L}_{11}\lvert_{y=0}=\mathcal{L}_7^{(n)}\, .
\end{equation}
We have verified by explicit computation that this relation indeed
holds for appropriate identifications of the truncated 11-dimensional fields
with the standard seven-dimensional fields which appear in Eq.~\eqref{7dsugra}.
For the generic $\mathbb{Z}_N$ orbifold
with $N>2$ and $n=1$, they are given by
\begin{eqnarray}
\si &=& \frac{3}{20}\ln\det g_{AB},  \label{id1}\\
\ti{g}_{\mu \nu } &=&e^{\frac{4}{3}\si }g_{\mu \nu }, \label{Weyl}\\
\psi_{\mu i} & = & \Psi_{\mu i}e^{\frac{1}{3}\si}-\frac{1}{5}\Up_\mu
 \left(\Gamma^A\Psi_{A}\right)_i e^{-\frac{1}{3}\si}, \\
\ti{C}_{\mu\nu\rho}&=&C_{\mu\nu\rho}, \\
\chi_i & = & \frac{3}{2\sqrt{5}}\left(\Gamma^A\Psi_{A}\right)_ie^{-\frac{1}{3}\si}, \label{idn}\\
F^I_{\mu\nu}&=&-\frac{i}{2}\mathrm{tr}\left( \si^IG_{\mu\nu}\right), \\
\la_i & =&  \frac{i}{2}\left( \Gamma^p\Psi_p - \Gamma^{\B{p}}\Psi_{\B{p}}\right)_ie^{-\frac{1}{3}\si}, \\
\ell_I^{\ph{I}\underline{J}}&=&\frac{1}{2}\mathrm{tr}\left( \B{\si}_I v \si^J v^{\dagger} \right).
 \label{idn2}
\ea
Furthermore the seven-dimensional supersymmetry spinor $\ve_i$ is related to
its 11-dimensional counterpart $\eta$ by
\begin{equation} \label{spinorrel}
\varepsilon_i=e^{\frac{1}{3}\si}\eta_i.
\end{equation}
In these identities, we have defined the matrices $G_{\mu\nu}\equiv
(G_{\mu\nu p\B{q}})$, $v\equiv
(e^{5\si/6}e^{\B{p}}_{\ph{p}\B{\underline{q}}})$ and made use of the
standard $SO(3,1)$ Pauli matrices $\si^I$, defined in Appendix
\ref{Pauli}.
For the $\mathbb{Z}_2$ orbifold we have $n=3$ and, therefore, two
additional $U(1)$ vector multiplets. Not surprisingly, field identification in
the gravity multiplet sector is unchanged from the generic case and still
given by Eqs.~\eqref{id1}-\eqref{idn}. It
is in the vector multiplet sector, where the additional states
appear, that we have to make a distinction. For the bosonic 
vector multiplet fields we find
\ba
F^I_{\mu\nu}&=&-\frac{1}{4}\mathrm{tr}\left( T^IG_{\mu\nu}\right),
\label{idn3} \\
\ell_I^{\ph{I}\underline{J}}&=&\frac{1}{4}\mathrm{tr}\left( \B{T}_I v
T^J v^{T} \right), \label{idn4} \ea where now $G_{\mu\nu}\equiv
(G_{\mu\nu AB})$ and $v\equiv
(e^{5\si/6}e^{A}_{\ph{A}\underline{B}})$. Here, $T^I$ are the six $SO(4)$
generators, which are explicitly given in Appendix \ref{Pauli}.


\section{General form of the supersymmetric bulk-brane action} \label{construct}
In this section, we present our general method of construction for the
full action, which combines 11-dimensional supergravity with the
seven-dimensional super-Yang-Mills theory on the orbifold plane in a
supersymmetric way. Main players in this construction will be the
constrained 11-dimensional bulk theory ${\cal L}_{11}$, as discussed
in Section~\ref{Bulk}, its truncation to seven dimensions, ${\cal
L}_7^{(n)}$, which has been discussed in the previous section and
corresponds to a $d=7$ EYM supergravity with gauge group $U(1)^n$ and
${\cal L}_{SU(N)}$, a $d=7$ EYM supergravity with gauge group
$U(1)^n\times SU(N)$. The $SU(N)$ gauge group in the latter theory
corresponds, of course, to the additional $SU(N)$ gauge multiplet
which one expects to arise for M-theory on the orbifold
$\mathbb{C}^2/\mathbb{Z}_N$.

Let us briefly discuss the physical origin of these $SU(N)$ gauge
fields on the orbifold fixed plane. It is well-known~\cite{Comp1,Lukas}, that
the $N-1$ Abelian $U(1)$ gauge fields within $SU(N)$ are already
massless for a smooth blow-up of the orbifold
$\mathbb{C}^2/\mathbb{Z}_N$ by an ALE manifold. More precisely, they
arise as zero modes of the M-theory three-form on the blow-up ALE
manifold. The remaining vector fields arise from membranes wrapping
the two-cycles of the ALE space and become massless only in the
singular orbifold limit, when these two-cycles collapse. For our
purposes, the only relevant fact is that all these $SU(N)$ vector fields are
located on the orbifold fixed plane. This allows us to treat the
Abelian and non-Abelian parts of $SU(N)$ on the same footing,
despite their different physical origin.\\

We claim that the action for the bulk-brane system is given by
\begin{equation} \label{action1}
S_{11-7}=\int_{\mathcal{M}^N_{11}} \mathrm{d}^{11}x \left[
 \mathcal{L}_{11} + \delta^{(4)}(y^A)\mathcal{L}_{\mathrm{brane}} \right],
\end{equation}
where
\begin{equation} \label{braneformula}
\mathcal{L}_{\mathrm{brane}}=\mathcal{L}_{\mathrm{SU(N)}}-\mathcal{L}_7^{(n)}.
\end{equation}
Here, as before, $\mathcal{L}_{11}$ is the Lagrangian for
11-dimensional supergravity~\eqref{11dsugra} with fields
constrained in accordance with the orbifold $\mathbb{Z}_N$ symmetry,
as discussed in Section~\ref{Bulk}. The Lagrangian
$\mathcal{L}_7^{(n)}$ describes a seven-dimensional ${\cal N}=1$ EYM
theory with gauge group $U(1)^n$. Choosing $n=1$ for generic
$\mathbb{Z}_N$ with $N>2$ and $n=3$ for $\mathbb{Z}_2$, this
Lagrangian corresponds to the truncation of the bulk Lagrangian
$\mathcal{L}_{11}$ to seven dimensions, as we have shown in the
previous section. This correspondence implies identifications between
truncated 11-dimensional bulk fields and the fields in
$\mathcal{L}_7^{(n)}$, which have also been worked out explicitly in
the previous section (see Eqs.~\eqref{id1}--\eqref{idn2} for the case
$N>2$ and Eqs.~\eqref{id1}--\eqref{idn} and \eqref{idn3}--\eqref{idn4}
for $N=2$).  These identifications are also considered part of the
definition of the Lagrangian~\eqref{action1}. The new Lagrangian
$\mathcal{L}_{SU(N)}$ is that of seven-dimensional EYM supergravity
with gauge group $U(1)^n\times SU(N)$, where, as usual, $n=1$ for
generic $\mathbb{Z}_N$ with $N>2$ and $n=3$ for $\mathbb{Z}_2$.
This Lagrangian contains the ``old'' states in the gravity multiplet
and the $U(1)^n$ gauge multiplet and the ``new'' states in the
$SU(N)$ gauge multiplet. We will think of the former states as being
identified with the truncated 11-dimensional bulk states by precisely
the same relations we have used for $\mathcal{L}_7^{(n)}$. The idea
of this construction is, of course, that in $\mathcal{L}_{\rm brane}$
the pure supergravity and $U(1)^n$ vector multiplet parts cancel
between ${\mathcal L}_{SU(N)}$  and $\mathcal{L}_7^{(n)}$, so that we
remain with ``pure'' $SU(N)$ theory on the orbifold plane. For this to
work out, we have to choose the seven-dimensional Newton constant
$\kappa_7$ within $\mathcal{L}_{SU(N)}$ to be the same as the one
in $\mathcal{L}_7^{(n)}$, that is
\begin{equation}
\kappa_7=\kappa^{5/9}\, .
\end{equation}
The supersymmetry transformation laws for the action~\eqref{action1}
are schematically given by
\ba \label{susy1}
\delta_{11}&=&\delta_{11}^{\ph{11}11}+\kappa^{8/9}\delta^{(4)}(y^A)\delta_{11}^{\ph{11}\mathrm{brane}}, \\
\delta_7&=&\delta_7^{\ph{7}\mathrm{SU(N)}}, \label{susy2}
\ea
where
\begin{equation} \label{susy3}
\delta_{11}^{\ph{11}\mathrm{brane}}=\delta_{11}^{\ph{11}\mathrm{SU(N)}}-\delta_{11}^{\ph{11}11}.
\end{equation}
Here $\delta_{11}$ and $\delta_7$ denote the supersymmetry transformation of
bulk fields and fields on the orbifold fixed plane, respectively.
A superscript 11 indicates a supersymmetry transformation
law of $\mathcal{L}_{11}$, as given in equations
\eqref{11dsusy1}, and a superscript $SU(N)$ indicates a
supersymmetry transformation law of $\mathcal{L}_{SU(N)}$, as can be
found by substituting the appropriate gauge group into equations
\eqref{7dsusy}. These transformation laws are parameterized by a
single 11-dimensional spinor, with the seven-dimensional spinors
in $\delta_{11}^{\ph{11}\mathrm{SU(N)}}$ and $\delta_{7}^{\ph{7}\mathrm{SU(N)}}$
being simply related to this 11-dimensional spinor by
equation \eqref{spinorrel}. On varying $S_{11-7}$ with respect to
these supersymmetry transformations we find
\ba
\delta S_{11-7}&=&-\int_{\mathcal{M}_{11}} \mathrm{d}^{11}x \delta^{(4)}(y^A)
\left( 1-\kappa^{8/9}\delta^{(4)}(y^A) \right) \delta_{11}^{\ph{11}\mathrm{brane}}
\mathcal{L}_{\mathrm{brane}}. \label{susycalc}
\ea
At first glance, the occurrence of delta-function squared terms is
concerning. However, as in Ho\v rava-Witten theory \cite{Hor-Wit}, we
can interpret the occurrence of these terms as a symptom of attempting
to treat in classical supergravity what really should be treated in
quantum M-theory. It is presumed that in proper quantum M-theory,
fields on the brane penetrate a finite thickness into the bulk, and that
there would be some kind of built-in cutoff allowing us to replace
$\delta^{(4)}(0)$ by a finite constant times $\kappa^{-8/9}$. If we
could set this constant to one and formally substitute
\begin{equation}
\delta^{(4)}(0)=\kappa^{-8/9}
\end{equation}
then the above integral would vanish.

As in Ref.~\cite{Hor-Wit}, we can avoid such a regularization if we
work only to lowest non-trivial order in $\kappa$, or, more precisely
to lowest non-trivial order in the parameter
$h=\kappa_7/g_\mathrm{YM}$. Note that $h$ has dimension of inverse
energy. To determine the order in $h$ of various terms in the
Lagrangian we need to fix a convention for the energy dimensions of
the fields. We assign energy dimension 0 to bulk bosonic fields and
energy dimension 1/2 to bulk fermions. This is consistent with the way
we have written down 11-dimensional supergravity \eqref{11dsugra}. In
terms of seven-dimensional supermultiplets this tells us to assign
energy dimension 0 to the gravity multiplet and the $U(1)$ vector
multiplet bosons and energy dimension 1/2 to the fermions in these
multiplets. For the $SU(N)$ vector multiplet, that is for the brane
fields, we assign energy dimension 1 to the bosons and 3/2 to the
fermions. With these conventions we can expand
\begin{equation} \label{expansion1}
\mathcal{L}_{SU(N)}=\kappa_7^{-2}\left( \mathcal{L}_{(0)}+h^2\mathcal{L}_{(2)}+
h^4\mathcal{L}_{(4)}+\ldots \right),
\end{equation}
where the $\mathcal{L}_{(m)}$, $m=0,2,4,\ldots$ are independent of $h$.
The first term in this
series is equal to $\mathcal{L}_7^{(n)}$, and therefore the leading
order contribution to $\mathcal{L}_{\mathrm{brane}}$ is precisely the
second term of order $h^2$ in the above series. It turns out that,
up to this order, the action $S_{11-7}$ is supersymmetric
under \eqref{susy1} and \eqref{susy2}. To see this we also expand
the supersymmetry transformation in orders of $h$, that is
\begin{equation}
\delta_{11}^{\ph{11}SU(N)} = \delta_{11}^{(0)}+h^2\delta_{11}^{(2)}+h^4\delta_{11}^{(4)}+\ldots\, .
\end{equation}
Using this expansion and Eq.~\eqref{expansion1} one finds that the
uncanceled variation \eqref{susycalc} is, in fact, of order
$h^4$. This means the action~\eqref{action1} is indeed supersymmetric
up to order $h^2$ and can be used to write down a supersymmetric
theory to this order. This is exactly what we will do in the following
section. We have also checked explicitly that the terms of order $h^4$
in Eq.~\eqref{susycalc} are non-vanishing, so that our method cannot
be extended straightforwardly to higher orders.

A final remark concerns the value of the Yang-Mills gauge coupling
$g_{\rm YM}$.  The above construction does not fix the value of this
coupling and our action is supersymmetric to order $h^2$ for all
values of $g_{\rm YM}$.  However, within M-theory one expects
$g_{\rm YM}$ to be fixed in terms of the 11-dimensional Newton constant
$\kappa$. Indeed, reducing M-theory on a circle to IIA string theory,
the orbifold seven-planes turn into D6 branes whose tension is fixed
in terms of the string tension~\cite{Polchinski}. By a straightforward
matching procedure this fixes the gauge coupling to be~\cite{Friedmann}
\begin{equation}
g_{\mathrm{YM}}^2=(4\pi)^{4/3}\kappa^{2/3}\; .
\end{equation}


\section{The explicit bulk/brane theory} \label{results}
In this section, we give a detailed description of M-theory on
$\mathcal{M}^N_{11}=\mathbb{R}^{1,6}\times\mathbb{C}^2/\mathbb{Z}_N$,
taking account of the additional states that appear on the brane. We begin
with a reminder of how the bulk fields, truncated to seven dimensions,
are identified with the fields that appear in the seven-dimensional
supergravity Lagrangians from which the theory is constructed.
Then we write down our full Lagrangian, and present the supersymmetry
transformation laws.\\

As discussed in the previous section, the full Lagrangian is
constructed from three parts, the Lagrangian of 11-dimensional
supergravity $\mathcal{L}_{11}$ with bulk fields constrained by the
orbifold action, $\mathcal{L}_7^{(n)}$, the Lagrangian of
seven-dimensional EYM supergravity with gauge group $U(1)^n$ and
$\mathcal{L}_{SU(N)}$, the Lagrangians for seven-dimensional EYM
supergravity with gauge group $U(1)^n\times SU(N)$. The Lagrangian
$\mathcal{L}_{11}$ has been written down and discussed in Section
\ref{Bulk}, whilst $\mathcal{L}_7^{(n)}$ has been dealt with in
Section \ref{reduction}. The final piece, $\mathcal{L}_{SU(N)}$, is
discussed in Appendix \ref{bigEYM}, where we provide the reader with a
general review of seven-dimensional supergravity theories. Crucial to
our construction is the way in which we identify the fields in the
supergravity and $U(1)^n$ gauge multiplets of the latter two
Lagrangians with the truncated bulk fields. Let us recall the
structure of this identification which has been worked out
in Section~\ref{reduction}. The bulk fields truncated to
seven dimensions form a $d=7$ gravity multiplet and $n$ $U(1)$
vector multiplets, where $n=1$ for the general $\mathbb{Z}_N$ orbifold
with $N>2$ and $n=3$ for the $\mathbb{Z}_2$ orbifold. The gravity multiplet
contains the purely seven-dimensional parts of the 11-dimensional
metric, gravitino and three-form, that is, $g_{\mu\nu}$, $\Psi_{\mu
i}$ and $C_{\mu\nu\rho}$, along with three vectors from $C_{\mu AB}$,
a spinor from $\Psi_{A\B{\imath}}$ and the scalar
$\mathrm{det}(e_A^{\ph{A}\underline{B}})$. Meanwhile, the vector
multiplets contain the remaining vectors from $C_{\mu AB}$, the
remaining spinors from $\Psi_{A\B{\imath}}$ and the scalars contained
in
$v_A^{\ph{A}\underline{B}}:=\mathrm{det}(e_A^{\ph{A}\underline{B}})^{-1/4}e_A^{\ph{A}\underline{B}}$,
the unit-determinant part of $e_A^{\ph{A}\underline{B}}$. The gravity
and $U(1)$ vector fields naturally combine together into a single
entity $A_\mu^I$, $I=1,\ldots(n+3)$, where the index $I$ transforms
tensorially under a global $SO(3,n)$ symmetry. Meanwhile, the vector
multiplet scalars naturally combine into a single $(3+n)\times (3+n)$
matrix $\ell$ which parameterizes the coset $SO(3,n)/SO(3)\times SO(n)$.
The precise mathematical form of these identifications is given in
equations \eqref{id1}-\eqref{idn2} for the general $\mathbb{Z}_N$
orbifold with $N>2$, and equations \eqref{id1}-\eqref{idn} and
\eqref{idn3}-\eqref{idn4} for the $\mathbb{Z}_2$ orbifold.

In addition to those states which arise from projecting bulk states to
the orbifold fixed plane the Lagrangian ${\cal L}_{SU(N)}$ also
contains a seven-dimensional $SU(N)$ vector multiplet, which is
genuinely located on the orbifold plane. It consists of gauge fields
$A_\mu^a$ with field strengths $F^a=\mathcal{D}A^a$, gauginos
$\la_i^a$, and $SU(2)$ triplets of scalars
$\phi_{a\ph{i}j}^{\ph{a}i}$. These fields are in the adjoint of
$SU(N)$ and we use $a,b,\ldots=4,\ldots,(N^2+2)$ for $su(N)$ Lie
algebra indices. It is important to write ${\cal L}_{SU(N)}$ in a form
where the $SU(N)$ states and the gravity/$U(1)^n$ states are
disentangled, since the latter must be identified with truncated bulk
states, as described above. For most of the fields appearing in
$\mathcal{L}_{SU(N)}$, this is just a trivial matter of using the
appropriate notation. For example, the vector fields in
$\mathcal{L}_{SU(N)}$ which naturally combine into a single entity
$A_\mu^{\ti{I}}$, where $\ti{I}=1,\ldots,(3+n+N^2-1)$, and transforms
as a vector under the global $SO(3,n+N^2-1)$ symmetry, can simply be
decomposed as $A_\mu^{\ti{I}}=(A_\mu^I,A_\mu^a)$, where $A_\mu^I$
refers to the three vector fields in the gravity multiplet and the
$U(1)^n$ vector fields and $A_\mu^a$ denotes the $SU(N)$ vector
fields. For gauge group $U(1)^n\times SU(N)$, the associated scalar
fields parameterize the coset $SO(3,n+N^2-1)/SO(3)\times SO(n+N^2-1)$.
We obtain representatives $L$ for this coset by expanding around
the bulk scalar coset $SO(3,n)/SO(3)\times SO(n)$, represented by matrices
$\ell$, to second order in the $SU(N)$ scalars $\Phi\equiv (\phi_a^{\ph{a}u})$.
For the details see Appendix~\ref{bigEYM2}. This leads to
\begin{equation}
L = \left( \begin{array}{ccc}
\ell+\frac{1}{2}h^2\ell\Phi^T\Phi & m & h\ell\Phi^T \\
h\Phi & 0 & \boldsymbol{1}_{N^2-1}+\frac{1}{2}h^2\Phi\Phi^T \\
\end{array} \right)\, .
\end{equation}
We note that the neglected $\Phi$ terms are of order $h^3$ and higher 
and, since we are aiming to construct the action only up to terms of order
$h^2$, are, therefore, not relevant in the present context. 
 
We are now ready to write down our final action.
As discussed in Section \ref{construct}, to order $h^2\sim g_{\mathrm{YM}}^{-2}$, it is given by
\begin{equation}
S_{11-7}=\int_{\mathcal{M}^N_{11}} \mathrm{d}^{11}x \left[ \mathcal{L}_{11} +
         \delta^{(4)}(y^A)\mathcal{L}_{\mathrm{brane}} \right],
\end{equation}
where
\begin{equation}
\mathcal{L}_{\mathrm{brane}}=\mathcal{L}_{\mathrm{SU(N)}}-\mathcal{L}_7^{(n)},
\end{equation}
and $n=3$ for the $\mathbb{Z}_2$ orbifold and $n=1$ for $\mathbb{Z}_N$
with $N>2$. The bulk contribution, $\mathcal{L}_{11}$, is given in
equation \eqref{11dsugra}, with bulk fields subject to the orbifold
constraints~\eqref{cond1}--\eqref{condn}. On the orbifold fixed plane,
$\mathcal{L}_7^{(n)}$ acts to cancel all the terms in
$\mathcal{L}_{\mathrm{SU(N)}}$ that only depend on bulk fields projected
to the orbifold plane. Thus none of the bulk gravity terms are replicated on the
orbifold place. To find $\mathcal{L}_{\mathrm{brane}}$ explicitly we need
to expand ${\cal L}_{SU(N)}$ in powers of $h$, using, in particular, the
above expressions for the gauge fields $A_\mu^{\tilde{I}}$ and the coset
matrices $L$, and extract the terms of order $h^2$. The further details 
of this calculation are provided in Appendix \ref{bigEYM}. The result is
\ba \label{braneaction}
\mathcal{L}_{\mathrm{brane}}\!\!\! &=&\!\!\! \frac{1}{g_{\mathrm{YM}}^2}
 \sqrt{-\ti{g}} \left\{ -\frac{1}{4}e^{-2\si}F^a_{\mu\nu}F_a^{\mu\nu}
-\frac{1}{2}\hat{\mathcal{D}}_\mu\phi_{a\ph{i}j}^{\ph{a}i}
\hat{\mathcal{D}}^\mu\phi_{\ph{aj}i}^{aj}-\frac{1}{2}\B{\la}^{ai}\Up^\mu\hat{\mathcal{D}}_\mu\la_{ai}
 -e^{-2\si}\ell_{I\ph{i}j}^{\ph{I}i}\phi_{a\ph{j}i}^{\ph{a}j}F^I_{\mu\nu}F^{a\mu\nu}\right.  \nn \\
&&\hspace{1.8cm} -\frac{1}{2}e^{-2\si}\ell_{I\ph{i}j}^{\ph{I}i}
{\phi^{aj}}_i\ell_{J\ph{k}l}^{\ph{J}k}\phi_{a\ph{l}k}^{\ph{a}l}
F^I_{\mu\nu}F^{J\mu\nu}-\frac{1}{2}p_{\mu\alpha\ph{i}j}^{\ph{\mu\alpha}i}
\phi_{a\ph{j}i}^{\ph{a}j}p^{\mu\alpha k}_{\ph{\mu\alpha k}l}\phi^{al}_{\ph{al}k} \nn \\
&&\hspace{1.8cm}+\frac{1}{4}\phi_{a\ph{i}k}^{\ph{a}i}
\hat{\mathcal{D}}_\mu\phi_{\ph{ak}j}^{ak}\left( \B{\psi}_\nu^j
\Up^{\nu\mu\rho}\psi_{\rho i} + \B{\chi}^j\Up^\mu\chi_i+ \B{\la}^{\alpha j}
\Up^\mu\la_{\alpha i} \right) \nn \\
&&\hspace{1.8cm}-\frac{1}{2\sqrt{2}}\left( \bar{\lambda}^{\alpha i}
\Upsilon ^{\mu \nu }\psi _{\mu j}+\bar{\lambda}^{\alpha i}\psi _{j}^{\nu }\right)
\phi_{a\ph{j}i}^{\ph{a}j}\phi^{ak}_{\ph{ak}l} p_{\nu\alpha \phantom{l}k}^{%
\phantom{\nu\alpha}l}-\frac{1}{\sqrt{2}}\left( \bar{\lambda}^{ai}
\Upsilon ^{\mu \nu }\psi_{\mu j}  +\bar{\lambda}^{ai}\psi_{j}^{\nu }\right)
\hat{\mathcal{D}}_\nu \phi_{ a\phantom{j}i}^{\phantom{ a}j} \nn 
\end{eqnarray}
\begin{eqnarray}
&& \hspace{1.8cm}+\frac{1}{192}e^{2\si }\ti{G}_{\mu \nu \rho
\sigma }\bar{\lambda}^{ai}\Upsilon ^{\mu
\nu \rho \sigma }\lambda_{ai}+\frac{i}{4\sqrt{2}}e^{-\si}F^I_{\mu\nu}
\ell_{I\ph{j}i}^{\ph{I}j}\B{\la}^{ai}\Up^{\mu\nu}\la_{aj}  \notag \\
&&\hspace{1.8cm}-\frac{i}{2}e^{-\si}\left(F_{\mu\nu}^I
\ell_{I\ph{k}l}^{\ph{I}k}\phi^{al}_{\ph{al}k}\phi^{\ph{a}j}_{a\ph{j}i}+
2F^a_{\mu\nu}\phi_{a\ph{j}i}^{\ph{a}j}\right)\left[ \frac{1}{4\sqrt{2}}\left( \bar{\psi}_{\rho
}^{i}\Upsilon ^{\mu \nu \rho \sigma }\psi _{\sigma j}+2\bar{\psi}^{\mu
i}\psi _{j}^{\nu }\right) \right.  \notag \\
&&\hspace{3.2cm}\left. +\frac{3}{20\sqrt{2}}\bar{\chi}^{i}\Upsilon ^{\mu \nu
}\chi _{j} -\frac{1}{4\sqrt{2}}\bar{\lambda}^{\alpha i}\Upsilon ^{\mu \nu
}\lambda _{\alpha j} +\frac{1}{2\sqrt{10}}\left( \bar{\chi}^{i}\Upsilon
^{\mu \nu \rho }\psi _{\rho j}-2\bar{\chi}^{i}\Upsilon ^{\mu }\psi _{j}^{\nu
}\right) \right]  \nn \\
&&\hspace{1.8cm} +e^{-\si}F_{a\mu\nu}\left[ \frac{1}{4}\left( 2\bar{\lambda}^{ai}
\Upsilon ^{\mu }\psi _{i}^{\nu }-\bar{%
\lambda}^{ai}\Upsilon ^{\mu \nu \rho }\psi _{\rho i}\right) +\frac{1}{2%
\sqrt{5}}\bar{\lambda}^{ai}\Upsilon ^{\mu \nu }\chi _{i}\right] \nn \\
&&\hspace{1.8cm}+\frac{1}{4}e^{2\si}f_{bc}^{\ph{bc}a}f_{dea}
\phi^{bi}_{\ph{bi}k}\phi^{ck}_{\ph{ck}j}\phi^{dj}_{\ph{dj}l}
\phi^{el}_{\ph{el}i} -\frac{1}{2}e^{\si}f_{abc}\phi^{bi}_{\ph{bi}k}\phi^{ck}_{\ph{ck}j}
\left(\bar{\psi}_{\mu}^{j}\Upsilon ^{\mu }\lambda_{i}^a +\frac{2}{\sqrt{5}}\bar{\chi}^{j}\lambda _{i}^{%
\phantom{i}a}\right) \notag \\
&&  \hspace{1.8cm}-\frac{i}{\sqrt{2}}e^{\si }f_{ab}^{\ph{ab}c}\phi_{c\ph{i}j}^{\ph{c}i}
\bar{\lambda}^{aj}\lambda _{i}^b+\frac{i}{60\sqrt{2}}e^{\si }f_{ab}^{\ph{ab}c}
\phi^{al}_{\ph{al}k}\phi^{bj}_{\ph{bj}l}\phi_{c\ph{k}j}^{\ph{c}k}\left(
5\bar{\psi}_{\mu }^{i}\Upsilon ^{\mu \nu }\psi _{\nu i}+2\sqrt{5}\bar{\psi}%
_{\mu }^{i}\Upsilon ^{\mu }\chi _{i}\right.  \notag \\
&& \hspace{5.7cm}\left.\left. +3\bar{\chi}^{i}\chi _{i}-5\B{\la}^{\alpha i}
\la_{\alpha i}\right)-\frac{1}{96}\epsilon ^{\mu \nu \rho \sigma \kappa \lambda
 \tau }\ti{C}_{\mu\nu \rho} F_{\si \kappa}^aF_{a\la \tau }\right\}.
\ea
Here $f_{ab}^{\ph{ab}c}$ are the structure constants of $SU(N)$.
The covariant derivatives that appear are given by
\ba
\mathcal{D}_\mu A_{\nu a}&=&\partial_\mu A_{\nu a} -\ti{\Gamma}^\rho_{\mu\nu}A_{\rho a}+
f_{ab}^{\phantom{ab}c}A_\mu^bA_\nu^c, \\
\hat{\mathcal{D}} _{\mu }\la_{ai}&=&\partial _{\mu }\la_{ai}+\frac{1}{2}
q_{\mu i}^{\phantom{\mu i}j}\la_{aj}+\frac{1}{4}\ti{\omega} _{\mu }^{\phantom{\mu}%
\underline{\mu }\underline{\nu }}\Upsilon _{\underline{\mu }\underline{\nu }%
}\la_{ai}+ f_{ab}^{\phantom{ab}c}A_\mu^b\la_{ci}, \\
\hat{\mathcal{D}}_\mu\phi_{a\ph{i}j}^{\ph{a}i}&=&\partial_\mu \phi_{a\ph{i}j}^{\ph{a}i}
 -q_{\mu \ph{i}j\ph{k}l}^{\ph{\mu}i\ph{j}k}\phi_{a\ph{l}k}^{\ph{a}l}+
f_{ab}^{\phantom{ab}c}A_\mu^b\phi_{c\ph{i}j}^{\ph{c}i},
\ea
with the Christoffel and spin connections $\ti{\Gamma}$ and
$\ti{\omega}$ taken in the seven-dimensional Einstein frame, (with
respect to the metric $\ti{g}$). Finally, the quantities $p$ and $q$
are the Maurer-Cartan forms of the bulk scalar coset matrix
$\ell_I^{\ph{I}\underline{J}}$ as given by equations
\eqref{Maurer1}--\eqref{Maurern}. Once again, the identities for
relating the seven-dimensional gravity and $U(1)$ vector multiplet fields to
11-dimensional bulk fields are given in equations
\eqref{id1}--\eqref{idn2} for the generic $\mathbb{Z}_N$ orbifold with $N>2$,
and equations \eqref{id1}--\eqref{idn} and \eqref{idn3}--\eqref{idn4} for
the $\mathbb{Z}_2$ orbifold. We stress that these identifications are
part of the definition of the theory.

The leading order brane corrections to the supersymmetry transformation laws
\eqref{11dsusy1} of the bulk fields are computed using equations
\eqref{susy1} and \eqref{susy2}. They are given by
\ba
\delta^{\mathrm{brane}}\psi_{\mu i}&=& \frac{\kappa_7^2}{g_{\mathrm{YM}}^2}
                                       \left\{ \frac{1}{2}\left( \phi_{ak}^{\ph{ak}j}
\hat{\mathcal{D}}_\mu\phi^{a\ph{i}k}_{\ph{a}i}-\phi^{a\ph{i}k}_{\ph{a}i}
\hat{\mathcal{D}}_\mu\phi_{ak}^{\ph{ak}j}\right)\ve_j -\frac{i}{15\sqrt{2}}
\Upsilon _{\mu }\varepsilon _{i}f_{ab}^{\ph{ab}c}\phi^{al}_{\ph{al}k}\phi^{bj}_{\ph{bj}l}
\phi_{c\ph{k}j}^{\ph{c}k}e^{\si} \right. \nn \\
&& \left. \hspace{1.1cm} +\frac{i}{10\sqrt{2}}\left(
\Upsilon _{\mu }^{\ph{\mu}\nu\rho }-8\delta _{\mu }^{\nu }\Upsilon ^{\rho
}\right) \varepsilon _{j}\left( F_{\nu\rho }^{I}\ell_{I\phantom{k}l}^{%
\phantom{I}k}\phi^{al}_{\ph{al}k}\phi^{\ph{a}j}_{a\ph{j}i}+2F^a_{\nu\rho}
\phi^{\ph{a}j}_{a\ph{j}i}\right)e^{-\si} \right\}, \nn \\
\delta^{\mathrm{brane}}\chi_i&=&\frac{\kappa_7^2}{g_{\mathrm{YM}}^2}
\left\{ -\frac{i}{2\sqrt{10}}\Up^{\mu\nu}\ve_j\left( F_{\mu\nu}^I
\ell_{I\ph{k}l}^{\ph{I}k}\phi^{al}_{\ph{al}k}\phi^{\ph{a}j}_{a\ph{j}i}+
2F^a_{\mu\nu}\phi^{\ph{a}j}_{a\ph{j}i}\right)e^{-\si} \right. \nn \\ 
&&\hspace{1.1cm}\left.+\frac{i}{3\sqrt{10}}\varepsilon _{i}f_{ab}^{\ph{ab}c}
\phi^{al}_{\ph{al}k}\phi^{bj}_{\ph{bj}l}\phi_{c\ph{k}j}^{\ph{c}k}e^{\si} \right\}, \nn \\
\ell_{I\ph{i}j}^{\ph{I}i}\delta^{\mathrm{brane}}A^I_\mu&=&
\frac{\kappa_7^2}{g_{\mathrm{YM}}^2}\left\{ \left( \frac{i}{\sqrt{2}}
 \B{\psi }_{\mu }^{k}\varepsilon _{l} -\frac{i}{%
\sqrt{10}}\B{\chi }^{k}\Upsilon _{\mu }\varepsilon _{l} \right)
\phi^{al}_{\ph{al}k}\phi^{\ph{a}i}_{a\ph{i}j} e^{\si}-\B{\ve}^k
\Up_\mu\la^a_k\phi_{a\ph{i}j}^{\ph{a}i}e^\si \right\}, \label{bulksusycorr}
\end{eqnarray}
\begin{eqnarray}
\ell_{I}^{\ph{I}\alpha}\delta^{\mathrm{brane}}A^I_\mu&=&0, \nn \\
\delta^{\mathrm{brane}}\ell_{I\ph{i}j}^{\ph{I}i}&=&
\frac{\kappa_7^2}{g_{\mathrm{YM}}^2}\left\{ \frac{i}{\sqrt{2}}
\left[ \B{\ve}^k\la_{\alpha l}\phi^{al}_{\ph{al}k}
\phi^{\ph{a}i}_{a\ph{i}j}\ell_I^{\ph{I}\alpha}+\B{\ve}^l
\la_{ak}\phi^{ai}_{\ph{ai}j}\ell_{I\ph{k}l}^{\ph{I}k}-
\left( \B{\ve}^i\la_{aj}-\frac{1}{2}\delta^i_j\B{\ve}^m\la_{am}\right)
 \phi^{al}_{\ph{al}k}\ell_{I\ph{k}l}^{\ph{I}k} \right] \right\},\nn \\
\delta^{\mathrm{brane}}\ell_{I}^{\ph{I}\alpha}&=&
\frac{\kappa_7^2}{g_{\mathrm{YM}}^2}\left\{ -\frac{i}{\sqrt{2}}\B{\ve}^i\la^\alpha_j
\phi^{aj}_{\ph{aj}i}\phi^{\ph{a}l}_{a\ph{l}k}\ell_{I\ph{k}l}^{\ph{I}k}\right\}, \nn \\
\delta^{\mathrm{brane}}\la_i^\alpha &=&\frac{\kappa_7^2}{g_{\mathrm{YM}}^2}
\left\{ \frac{i}{\sqrt{2}}\Up^\mu\ve_j\phi_{ai}^{\ph{ai}j}
p_{\mu \ph{\alpha k}l}^{\ph{\mu}\alpha k}\phi^{al}_{\ph{al}k}\right\}, \nn
\ea
where $\ve_i$ is the 11-dimensional supersymmetry spinor $\eta$
projected onto the orbifold plane, as given in \eqref{spinorrel}. We note that
not all of the bulk fields receive corrections to their
supersymmetry transformation laws. The leading order supersymmetry
transformation laws of the $SU(N)$ multiplet fields are found using equation
\eqref{susy2} and take the form
\ba
\delta A_\mu^a&=& \B{\ve}^i\Up_\mu\la_i^ae^\si-\left( i\sqrt{2}\psi_\mu^i\ve_j
 -\frac{2i}{\sqrt{10}}\B{\chi}^i\Up_\mu\ve_j \right)\phi^{aj}_{\ph{aj}i}e^\si,\nn \\
\delta \phi_{a\ph{i}j}^{\ph{a}i}&=&-i\sqrt{2}\left( \B{\ve}^i\la_{aj}-
\frac{1}{2}\delta^i_j\B{\ve}^k\la_{ak}\right), \label{susybranexmfn} \\
\delta\la^a_i&=&-\frac{1}{2}\Up^{\mu\nu}\ve_i\left( F^I_{\mu\nu}
\ell^{\ph{I}j}_{I\ph{j}k}\phi^{ak}_{\ph{ak}j}+F^a_{\mu\nu}\right)e^{-\si}-i\sqrt{2}
\Up^\mu\ve_j\hat{\mathcal{D}}_\mu\phi^{a\ph{i}j}_{\ph{a}i}-i\ve_j
f^a_{\ph{a}bc}\phi^{bj}_{\ph{bj}k}\phi^{ck}_{\ph{ck}i}. \nn 
\ea
\\

To make some of the properties of our result more transparent, it is
helpful to extract the bosonic part of the action. This bosonic part
will also be sufficient for many practical applications. We recall
that the full Lagrangian \eqref{braneaction} is written in the
seven-dimensional Einstein frame to avoid the appearance of
$\sigma$--dependent pre-factors in many terms. The bosonic part,
however, can be conveniently formulated in terms of $g_{\mu\nu}$, the
seven-dimensional part of the 11-dimensional bulk metric
$g_{MN}$. This requires performing the Weyl-rescaling~\eqref{Weyl}.
It also simplifies the notation if we rescale the scalar $\si$ as
$\tau=10\si/3$, and drop the tilde from the three-form
$\tilde{C}_{\mu\nu\rho}$ and its field strength
$\tilde{G}_{\mu\nu\rho\sigma}$, which exactly coincide with the purely
seven-dimensional components of their 11-dimensional counterparts.
Let us now write down the purely bosonic part of our action, subject
to these small modifications. We find
\begin{equation}
\mathcal{S}_{11-7,{\rm bos}}=\mathcal{S}_{11,{\rm bos}} + \mathcal{S}_{7,{\rm bos}}\, ,
\end{equation}
where $\mathcal{S}_{11,{\rm bos}}$ is the bosonic part of 11-dimensional
supergravity~\eqref{11dsugra}, with fields subject to the orbifold constraints
\eqref{cond1}--\eqref{cond6}. Further, $\mathcal{S}_{7,{\rm bos}}$ is the bosonic
part of Eq.~\eqref{braneaction}, subject to the above modifications, for which
we obtain
\ba \label{boseaction}
\mathcal{S}_{7,{\rm bos}}&=& \frac{1}{g_{\mathrm{YM}}^2} \int_{y=0} \mathrm{d}^7x \sqrt{-g}
\left( -\frac{1}{4}H_{ab}F^a_{\mu\nu}F^{b\mu\nu} -\frac{1}{2}H_{aI}F^a_{\mu\nu}
F^{I\mu\nu} -\frac{1}{4}(\delta H)_{IJ}F^I_{\mu\nu}F^{J\mu\nu}  \right. \nn \\
&&\hspace{3.5cm} \left. -\frac{1}{2}e^\tau\hat{\mathcal{D}}_\mu
\phi_{a\ph{i}j}^{\ph{a}i}\hat{\mathcal{D}}^\mu\phi_{\ph{aj}i}^{aj}
-\frac{1}{2}\left( \delta K \right)^{\alpha j \ph{i} \beta l}_{\ph{\alpha j}i
\ph{\beta l} k}p_{\mu \alpha \ph{i} j}^{\ph{\mu\alpha}i\ph{j}}
p^{\mu \ph{\beta}k}_{\ph{\mu}\beta\ph{k}l}+\frac{1}{4}D^{ai}_{\ph{ai}j}D^{\ph{a}j}_{a\ph{j}i}\right) \nn \\
&&-\frac{1}{4g_{\mathrm{YM}}^2}\int_{y=0} C\wedge F^a \wedge F_a,
\ea
where
\ba
H_{ab}&=&\delta_{ab}, \label{H1} \\
H_{aI}&=&2\ell_{I\ph{i}j}^{\ph{I}i}\phi_{a\ph{j}i}^{\ph{a}j}, \label{H2} \\
(\delta H)_{IJ}&=&2\ell_{I\ph{i}j}^{\ph{I}i}\phi_{a\ph{j}i}^{\ph{a}j}
\ell_{J\ph{k}l}^{\ph{J}k}\phi_{a\ph{l}k}^{\ph{a}l},  \label{H3} \\
\left( \delta K \right)^{\alpha j \ph{i} \beta l}_{\ph{\alpha j}i\ph{\beta l} k}&=&
 e^\tau \delta^{\alpha\beta} \phi_{a\ph{j}i}^{\ph{a}j}\phi^{al}_{\ph{al}k}, \\
D^{ai}_{\ph{ai}j}&=&e^\tau f^a_{\ph{a}bc}\phi^{bi}_{\ph{bi}k}\phi^{ck}_{\ph{ck}j}.
\ea
The gauge covariant derivative is denoted by $\mathcal{D}$, whilst $\hat{\mathcal{D}}$ is given by
\begin{equation}
\hat{\mathcal{D}}_\mu\phi_{a\ph{i}j}^{\ph{a}i}=\mathcal{D}_\mu
 \phi_{a\ph{i}j}^{\ph{a}i} -q_{\mu \ph{i}j\ph{k}l}^{\ph{\mu}i\ph{j}k}\phi_{a\ph{l}k}^{\ph{a}l}\, .
\end{equation}
The Maurer-Cartan forms $p$ and $q$ of the matrix of scalars $\ell$ are defined by
\ba 
p_{\mu\alpha \ph{i}j}^{\ph{\mu\alpha }i}&=&\ell^{I}_{\ph{I}\alpha}\partial_\mu
 \ell_{I\ph{i}j}^{\ph{\mu}i}, \\
q_{\mu \ph{i}j\ph{k}l}^{\ph{\mu}i\ph{j}k}&=&\ell^{Ii}_{\ph{Ii}j}\partial_\mu
 \ell_{I\ph{k}l}^{\ph{\mu}k}.
\ea
The bosonic fields localized on the orbifold plane are the $SU(N)$
gauge vectors $F^a=\mathcal{D}A^a$ and the $SU(2)$ triplets of scalars
$\phi_{a\ph{i}j}^{\ph{a}i}$. All other fields are projected from the
bulk onto the orbifold plane, and there are algebraic equations relating
them to the 11-dimensional fields in $\mathcal{S}_{11}$.
As discussed above, these relations are trivial for the metric $g_{\mu\nu}$
and the three-form $C_{\mu\nu\rho}$, whilst the scalar $\tau$ is given by
\begin{equation}
\tau = \frac{1}{2}\ln\det g_{AB}\, ,
\end{equation}
and can be interpreted as an overall scale factor of the orbifold
$\mathbb{C}^2/\mathbb{Z}_N$. For the remaining fields, the ``gravi-photons''
$F^I_{\mu\nu}$ and the ``orbifold moduli'' $\ell_I^{\ph{I}\underline{J}}$,
we have to distinguish between the generic $\mathbb{Z}_N$ orbifold with
$N>2$ and the $\mathbb{Z}_2$
orbifold. For $\mathbb{Z}_N$ with $N>2$ we have four $U(1)$ gauge fields,
so that $I=1,\ldots ,4$, and $\ell_I^{\ph{I}\underline{J}}$ parameterizes
the coset $SO(3,1)/SO(3)$. They are identified with 11-dimensional fields through
\ba
F^I_{\mu\nu}&=&-\frac{i}{2}\mathrm{tr}\left( \si^IG_{\mu\nu}\right), \\
\ell_I^{\ph{I}\underline{J}}&=&\frac{1}{2}\mathrm{tr}\left( \B{\si}_I v \si^J v^{\dagger} \right),
\ea
where $G_{\mu\nu}\equiv (G_{\mu\nu p\B{q}})$,
$v\equiv (e^{\tau/4}e^{\B{p}}_{\ph{p}\B{\underline{q}}})$ and $\si^I$
are the $SO(3,1)$ Pauli matrices as given in Appendix \ref{Pauli}.
For the $\mathbb{Z}_2$ case, we have six $U(1)$ vector fields, so that
$I=1,\ldots ,6$, and $\ell_I^{\ph{I}\underline{J}}$ parameterizes the
coset $SO(3,3)/SO(3)^2$. The field identifications now read
\ba
F^I_{\mu\nu}&=&-\frac{1}{4}\mathrm{tr}\left( T^IG_{\mu\nu}\right), \\
\ell_I^{\ph{I}\underline{J}}&=&\frac{1}{4}\mathrm{tr}\left( \B{T}_I v T^J v^{T} \right),
\ea
where this time $G_{\mu\nu}\equiv (G_{\mu\nu AB})$,
$v\equiv (e^{\tau/4}e^{A}_{\ph{A}\underline{B}})$, and $T^I$
are the generators of $SO(4)$, as given in Appendix \ref{Pauli}.\\

Let us discuss a few elementary properties of the bosonic
action~\eqref{boseaction} on the orbifold plane, starting with the
gauge-kinetic functions \eqref{H1}--\eqref{H3}. The first observation
is, that the gauge-kinetic function for the $SU(N)$ vector fields is
trivial (to the order we have calculated), which confirms the result
of Ref.~\cite{Friedmann}. On the other hand, we find non-trivial gauge
kinetic terms between the $SU(N)$ vectors and the gravi-photons, as
well as between the gravi-photons. We also note the appearance of the
Chern-Simons term $C\wedge F^a\wedge F_a$, which has been
predicted~\cite{Anomaly} from anomaly cancellation in configurations
which involve additional matter fields on conical singularities, but,
in our case, simply follows from the structure of seven-dimensional
supergravity without any further assumption. We note that, while there
is no seven-dimensional scalar field term which depends only on
orbifold moduli, the scalar field kinetic terms in~\eqref{boseaction}
constitute a complicated sigma model which mixes the orbifold moduli
and the scalars in the $SU(N)$ vector multiplets. A further interesting
feature is the presence of the seven-dimensional D-term potential in
Eq.~\eqref{boseaction}. Introducing the matrices $\phi_a\equiv
(\phi^{\ph{a}i}_{a\ph{i}j})$ and $D^a\equiv (D^{ai}_{\ph{ai}j})$
this potential can be written as
\begin{equation}
 V=\frac{1}{4g_{\rm YM}^2}{\rm tr}\left( D^aD_a\right)\, ,
\end{equation}
where
\begin{equation}
D^a=\frac{1}{2}e^\tau f^a_{\ph{a}bc}[\phi^b,\phi^c]\; .
\end{equation}
The flat directions, $D^a=0$, of this potential, which correspond to
unbroken supersymmetry as can be seen from Eq.~\eqref{susybranexmfn},
can be written as
\begin{equation}
\phi^a=v^a\sigma^3
\end{equation}
with vacuum expectation values $v^a$. The $v^a$ correspond to elements
in the Lie algebra of $SU(N)$ which can be diagonalised into the
Cartan sub-algebra. Generic such diagonal matrices break $SU(N)$
to $U(1)^{N-1}$, while larger unbroken groups are possible for non-generic
choices. Looking at the scalar field masses induced from the D-term in such a
generic situation, we have one massless scalar for each
of the non-Abelian gauge fields which is absorbed as their longitudinal
degree of freedom. For each of the $N-1$ unbroken Abelian gauge fields,
we have all three associated scalars massless, as must be the case from
supersymmetry. This situation corresponds exactly to what happens when
the orbifold singularity is blown up. We can, therefore, see that within
our supergravity construction blowing-up is encoded by the D-term.
Further, the Abelian gauge fields in $SU(N)$ correspond to (a truncated version of)
the massless vector fields which arise from zero modes of the M-theory three-form
on a blown-up orbifold, while the $3(N-1)$ scalars in the Abelian vector fields
correspond to the blow-up moduli.


\section{Discussion and outlook}

In this paper, we have constructed the effective supergravity action for
M-theory on the orbifold $\mathbb{C}^2/\mathbb{Z}_N\times\mathbb{R}^{1,6}$,
by coupling 11-dimensional supergravity, constrained in accordance with
the orbifolding, to $SU(N)$ super-Yang-Mills theory located on the
seven-dimensional fixed plane of the orbifold. We have found that the
orbifold-constrained fields of 11-dimensional supergravity, when restricted
to the orbifold plane, fill out a seven-dimensional supergravity multiplet
plus a single $U(1)$ vector multiplet for $N>2$ and three $U(1)$ vector
multiplets for $N=2$. The seven-dimensional action on the orbifold plane,
which has to be added to 11-dimensional supergravity, couples these bulk
degrees of freedom to genuine seven-dimensional states in the $SU(N)$
multiplet. We have obtained this action on the orbifold plane by ``up-lifting''
information from the known action of ${\cal N}=1$ Einstein-Yang-Mills
supergravity and identifying 11- and 7-dimensional degrees of freedom appropriately.
The resulting 11-/7-dimensional theory is given as an expansion in the
parameter $h=\kappa^{5/9}/g_{\rm YM}$, where $\kappa$ is the 11-dimensional
Newton constant and $g_{\rm YM}$ is the seven-dimensional $SU(N)$ coupling.
The bulk theory appears at zeroth order in $h$, and we have determined the
complete set of leading terms on the orbifold plane which
are of order $h^2$. At order $h^4$ we encounter a singularity due to
a delta function square, similar to what happens in Ho\v{r}ava-Witten
theory~\cite{Hor-Wit}. As in Ref.~\cite{Hor-Wit}, we assume that this singularity
will be resolved in full M-theory, when the finite thickness of the
orbifold plane is taken into account, and that it does not invalidate the
results at order $h^2$. 

While we have focused on the A-type orbifolds
$\mathbb{C}^2/\mathbb{Z}_N$, we expect our construction to work
analogously for the other four-dimensional orbifolds of ADE type. Our
result represents the proper starting point for compactifications of
M-theory on $G_2$ spaces with singularities of the type
$\mathbb{C}^2/\mathbb{Z}_N\times B$, where $B$ is a three-dimensional
manifold. We consider this to be the first step in a programme, aiming
at developing an explicit supergravity framework for ``phenomenological''
compactifications of M-theory on singular $G_2$ spaces. A further
important step would be to couple to our action a four-dimensional
${\cal N}=1$ action, describing the matter fields on conical singularities.
The structure of such a coupled supergravity in eleven, seven and four
dimensions is currently under investigation.

\section*{Acknowledgments}
L.~B.~A.~is supported by an NSF Graduate Research Fellowship,
A.~B.~B.~is supported by a PPARC Postgraduate Studentship and A.~L.~is
supported by a PPARC Advanced Fellowship.

\section*{Appendix}

\appendix        

\section{Spinor Conventions} \label{spinors} 
In this section, we provide the conventions for gamma matrices and
spinors in eleven, seven and four dimensions and the relations between
them. This split of eleven dimensions into seven plus four arises
naturally from the orbifolds
$\mathbb{R}^{1,6}\times\mathbb{C}^{2}/\mathbb{Z}_{N}$ which we
consider in this paper. We need to work out the appropriate spinor
decomposition for this product space and, in particular, write
11-dimensional Majorana spinors as a product of seven-dimensional
symplectic Majorana spinors with an appropriate basis of
four-dimensional spinors. We denote 11-dimensional coordinates by
$(x^M)$, with indices $M,N,\ldots = 0,\ldots , 10$. They are split up
as $x^{M}=(x^{\mu},y^{A})$ with seven-dimensional coordinates $x^\mu$,
where $\mu ,\nu ,\ldots = 0,\ldots ,6$, on $\mathbb{R}^{1,6}$ and
four-dimensional coordinates $y^A$, where $A,B,\ldots = 7,\ldots ,10$, on
$\mathbb{C}^2/\mathbb{Z}_N$.\\

We begin with gamma matrices and spinors in 11-dimensions. The gamma-matrices,
$\Gamma^{M}$, satisfy the standard Clifford algebra
\begin{equation}
\{\Gamma^{M},\Gamma^{N}\}=2g^{MN},\label{clifford}%
\end{equation}
where $g^{MN}$ is the metric on the full space $\mathbb{R}^{1,6}%
\times\mathbb{C}^{2}/\mathbb{Z}_{N}$. We define the Dirac conjugate of an
11-dimensional spinor $\Psi$ to be
\begin{equation}
\bar{\Psi}=i\Psi^{\dagger}\Gamma^{0}.
\end{equation}
The 11-dimensional charge conjugate is given by
\begin{equation}
\Psi^{C}=B^{-1}\Psi^{\ast},
\end{equation}
where the charge conjugation matrix $B$ satisfies \cite{Tanii}
\begin{equation}
B\Gamma^{M}B^{-1}=\Gamma^{M\ast},\hspace{0.5cm}B^{\ast}B=\boldsymbol{1} _{32}.
\end{equation}
In this work, all spinor fields in 11-dimensions are taken to satisfy the
Majorana condition, $\Psi^{C}=\Psi$, thereby reducing $\Psi$ from 32 complex
to 32 real degrees of freedom.\\

Next, we define the necessary conventions for $SO(1,6)$ gamma matrices and
spinors in seven dimensions. The gamma matrices, denoted by $\Upsilon^\mu$,
satisfy the algebra
\begin{equation}
 \{\Upsilon^\mu ,\Upsilon^\nu\}=2g^{\mu\nu}\; ,
\end{equation}
where $g_{\mu\nu}$ is the metric on $\mathbb{R}^{1,6}$. The Dirac conjugate
of a general eight complex component spinor $\psi$ is defined by
\begin{equation}
\bar{\psi}=i\psi^{\dagger}\Upsilon^{0}\; .
\end{equation}
In seven dimensions, the charge conjugation matrix $B_8$ has the
following properties \cite{Tanii}
\begin{equation}
B_{8}\Upsilon^{\mu}B_{8}^{-1}=\Upsilon^{\mu\ast},\hspace{0.5cm}B_{8}^{\ast
}B_{8}=-\boldsymbol{1} _{8}\; .
\end{equation}
The second of these relations implies that charge conjugation, defined by
\begin{equation}
 \psi^c = B_8^{-1}\psi^\ast
\end{equation}
squares to minus one. Hence, one cannot define seven-dimensional
$SO(1,6)$ Majorana spinors.  However, the supersymmetry algebra in seven
dimensions contains an $SU(2)$ R-symmetry and spinors can be naturally
assembled into $SU(2)$ doublets $\psi^i$, where $i,j,\ldots =
1,2$. Indices $i,j,\ldots$ can be lowered and raised with the
two-dimensional Levi-Civita tensor $\epsilon_{ij}$ and $\epsilon^{ij}$,
normalized so that $\epsilon^{12}=\epsilon_{21}=1$. With these conventions
a symplectic Majorana condition
\begin{equation}
\psi_{i}=\epsilon_{ij}B_{8}^{-1}\psi^{\ast j},\label{symplectic majorana}%
\end{equation}
can be imposed on an $SU(2)$ doublet $\psi^i$ of spinors, where we have
defined $\psi^{\ast i}\equiv(\psi_{i})^{\ast}$. All seven-dimensional
spinors in this paper are taken to be such symplectic Majorana spinors.
Further, in computations with seven-dimensional spinors, the following
identities are frequently useful,
\begin{align}
\bar{\chi}^{i}\Upsilon^{\mu_{1}\ldots\mu_{n}}\psi^{j} &  =(-1)^{n+1}\bar{\psi
}^{j}\Upsilon^{\mu_{n}\ldots\mu_{1}}\chi^{i},\\
\bar{\chi}^{i}\Upsilon^{\mu_{1}\ldots\mu_{n}}\psi_{i} &  =(-1)^{n}\bar{\psi
}^{i}\Upsilon^{\mu_{n}\ldots\mu_{1}}\chi_{i}.
\end{align}
\\

Finally, we need to fix conventions for four-dimensional Euclidean gamma
matrices and spinors. Four-dimensional gamma matrices, denoted by
$\gamma^A$, satisfy
\begin{equation}
 \{\gamma^A,\gamma^B\}=2g^{AB}\, ,
\end{equation}
with the metric $g_{AB}$ on $\mathbb{C}^2/\mathbb{Z}_N$. The chirality
operator, defined by 
\begin{equation}
\gamma=\gamma^{\underline{7}}\gamma^{\underline{8}}\gamma^{\underline{9}}
       \gamma^{\underline{10}}\, ,
\end{equation}
satisfies $\gamma^2 = {\bf 1}_4$. The four-dimensional charge conjugation
matrix $B_4$ satisfies the properties
\begin{equation}
B_4\gamma^AB_4^{-1}=\gamma^{A\ast}\; ,\qquad B_4^\ast B_4=-{\bf 1}_4\; .\label{B4}
\end{equation}
It will often be more convenient to work with complex coordinates
$(z^p,\bar{z}^{\bar{p}})$ on $\mathbb{C}^2/\mathbb{Z}_N$, where
$p,q,\ldots = 1,2$ and $\bar{p},\bar{q},\ldots = \bar{1},\bar{2}$.
In these coordinates, the Clifford algebra takes the well-known
``harmonic oscillator'' form
\begin{equation}
\left\{  \gamma^{p},\gamma^{q}\right\}  =0\;, \qquad
\left\{  \gamma^{\bar{p}},\gamma^{\bar{q}}\right\}  =0\; ,\qquad
\left\{  \gamma^{p},\gamma^{\bar{q}}\right\}     =2g^{p\bar{q}}\, ,
\end{equation}
with creation and annihilation ``operators'' $\gamma^p$ and $\gamma^{\bar{p}}$,
respectively. In this new basis, complex conjugation of gamma matrices~\eqref{B4}
is described by 
\begin{equation}
B_{4}\gamma^{\bar{p}}B_{4}^{-1}=\gamma^{p\ast}\; , \qquad B_{4}\gamma
^{p}B_{4}^{-1}=\gamma^{\bar{p}\ast}\; . \label{B4c}
\end{equation}
 A basis of spinors can be obtained by starting with the
``vacuum state'' $\Omega$, which is annihilated by $\gamma^{\bar{p}}$, that
is $\gamma^{\bar{p}}\Omega =0$, and applying creation operators to it.
This leads to the three further states 
\begin{equation}
\rho^{\underline{p}}=\frac{1}{\sqrt{2}}\gamma^{\underline{p}}\Omega\; ,\qquad
\bar{\Omega}=\frac{1}{2}\gamma^{\ul{1}}\gamma^{\ul{2}}\Omega\, .
\end{equation}
In terms of the gamma matrices in complex coordinates, the chirality operator
$\gamma$ can be expressed as
\begin{equation}
\gamma=-1+\gamma^{\bar{\ul 1}}\gamma^{\ul 1}+\gamma^{\bar{\ul 2}}\gamma^{\ul 2}-\gamma
^{\bar{\ul 1}}\gamma^{\ul 1}\gamma^{\bar{\ul 2}}\gamma^{\ul 2}\, .
\end{equation}
Hence, the basis $(\Omega ,\rho^{\ul p},\bar{\Omega})$ consists of chirality
eigenstates satisfying
\begin{equation}
\gamma\Omega=-\Omega\; , \qquad \gamma\bar{\Omega}=-\bar{\Omega}\; , \qquad
\gamma\rho^{\underline{p}} = \rho^{\underline{p}}\; . 
\end{equation}
For ease of notation, we will write the left-handed states as
$(\rho^i )=(\rho^{\ul{1}},\rho^{\ul{2}})$, where $i ,j ,\ldots =1,2$ 
and the right-handed states as $(\rho^{\bar{\imath}})=(\Omega ,\bar{\Omega})$
where $\bar{\imath},\bar{\jmath},\ldots =\bar{1},\bar{2}$. Note, it follows
from Eq.~\eqref{B4c} that
\begin{equation}
 B_4^{-1}\Omega^\ast = \bar{\Omega}\; ,\qquad
 B_4^{-1}\rho^{\ul{1}\ast} = \rho^{\ul{2}}\; .
\end{equation}
Hence $\rho^i$ and $\rho^{\bar{\imath}}$ each form a Majorana 
conjugate pair of spinors with definite chirality.\\

We should now discuss the four plus seven split of 11-dimensional gamma
matrices and spinors. It is easily verified that the matrices
\begin{equation}
\Gamma^{\mu}    =\Upsilon^{\mu}\otimes\gamma\; ,\qquad
\Gamma^{A}    =\boldsymbol{1}  _{8}\otimes\gamma^{A}, \label{gammas}
\end{equation}
satisfy the Clifford algebra~\eqref{clifford} and, hence, constitute a
valid set of 11-dimensional gamma-matrices. Further, it is clear that
an 11-dimensional charge conjugation matrix $B$ can be obtained from
its seven- and four-dimensional counterparts $B_8$ and $B_4$ by
\begin{equation}
 B=B_8\otimes B_4\; .
\end{equation}
A general 11-dimensional Dirac spinor $\Psi$ can now be expanded in terms
of the basis $(\rho^i,\rho^{\bar{\imath}})$ of four-dimensional spinors
as
\begin{equation}
\Psi =\psi_{i}(x,y)\otimes\rho^{i}+\psi_{\bar{\jmath}}(x,y)\otimes\rho
^{\bar{\jmath}}\; ,
\end{equation}
where $\psi_i$ and $\psi_{\bar{\jmath}}$ are four independent seven-dimensional
Dirac spinors. Given the properties of the four-dimensional spinor basis under
charge conjugation, a Majorana condition on the 11-dimensional spinor $\Psi$
simply translates into $\psi_i$ and $\psi_{\bar{\jmath}}$ each being
symplectic $SO(1,6)$ Majorana spinors.

\section{Some group-theoretical properties}
\label{Pauli} 
In this section we summarize some group-theoretical properties related
to the coset spaces $SO(3,n)/$ $SO(3)\times SO(n)$ of seven-dimensional
EYM supergravity. We focus on the parameterization of these coset
spaces in terms of 11-dimensional metric components, which is an
essential ingredient in re-writing 11-dimensional supergravity,
truncated on the orbifold, into standard seven-dimensional EYM
supergravity language.

We begin with the generic $\mathbb{C}^2/\mathbb{Z}_N$ orbifold, where
$N>2$ and $n=1$, so the relevant coset space is $SO(3,1)/SO(3)$. In
this case, it is convenient to use complex coordinates
$(z^p,\bar{z}^{\bar{p}})$, where $p,q,\ldots = 1,2$ and
$\bar{p},\bar{q},\ldots = \bar{1},\bar{2}$, on the orbifold.  After
truncating the 11-dimensional metric to be independent of the orbifold
coordinates, the surviving degrees of freedom of the orbifold part of
the metric can be described by the components ${e_p}^{\ul{p}}$ of the
vierbein, see Eqs.~\eqref{cond1}--\eqref{condn}. Extracting the
overall scale factor from this, we have a determinant one object
${v_p}^{\ul{p}}$, together with identifications by $SU(2)$ gauge
transformations acting on the tangent space index. Hence, ${v_p}^{\ul{p}}$ should
be thought of as parameterizing the coset $SL(2,\mathbb{C})/SU(2)$. This
space is indeed isomorphic to $SO(3,1)/SO(3)$. To work this out explicitly,
it is useful to introduce the map $f$ defined by
\begin{equation}
 f(u) = u_I\sigma^I
\end{equation}
which maps four-vectors $u_I$, where $I,J,\ldots =1,\ldots ,4$,
into hermitian matrices $f(u)$. Here the matrices $\sigma^I$ and their
conjugates $\bar{\sigma}^I$ are given by
\begin{equation}
(\sigma^{I})=(\sigma^{u},\boldsymbol{1}  _{2})\; ,\qquad
(\bar{\sigma}^{I})=(-\sigma^{u},\boldsymbol{1}  _{2})\; ,
\end{equation}
where the $\sigma^{u}$, $u=1,2,3$, are the standard Pauli matrices.
They satisfy the following useful identities
\begin{align}
\mathrm{tr}\left(  \sigma^{I}\bar{\sigma}^{J}\right)   &  =2\eta^{IJ},\\
\mathrm{tr}\left(  \bar{\sigma}^{I}\sigma^{(J}\bar{\sigma}^{\lvert K\rvert
}\sigma^{L)}\right)   &  =2\left(  \eta^{IJ}\eta^{KL}+\eta^{IL}\eta^{JK}%
-\eta^{IK}\eta^{JL}\right)\,  ,
\end{align}
where $I,J,\ldots$ indices are raised and lowered with the Minkowski metric
$(\eta_{IJ})=\rm{diag}(-1,-1,-1,+1)$. A key property of the map $f$ is that
\begin{equation}
 u_Iu^I={\rm det}(f(u))
\end{equation}
for four-vectors $u_I$. This property is crucial in demonstrating that
the map $F$ defined by
\begin{equation}
 F(v)u = f^{-1}\left( vf(u)v^\dagger\right)
\end{equation}
is a group homeomorphism $F:SL(2,\mathbb{C})\rightarrow SO(3,1)$.
Solving explicitly for the $SO(3,1)$ images ${\ell_I}^{\ul{J}}={(F(v))_I}^{\ul{J}}$
one finds
\begin{equation}
 {\ell_I}^{\ul{J}}=\frac{1}{2}{\rm tr}\left(\bar{\sigma}_Iv\sigma^Jv^\dagger\right)\; .
\end{equation}
This map induces the desired map $SL(2,\mathbb{C})/SU(2)\rightarrow SO(3,1)/SO(3)$
between the cosets.\\

The structure is analogous, although slightly more involved, for the orbifold
$\mathbb{C}^2/\mathbb{Z}_2$, where $n=3$ and the relevant coset space is
$SO(3,3)/SO(3)^2$. In this case, it is more appropriate to work with real
coordinates $y^A$ on the orbifold, where $A,B,\ldots =7,8,9,10$. The
orbifold part of the truncated 11-dimensional metric, rescaled to
determinant one, is then described by the vierbein ${v_A}^{\ul{A}}$ in real
coordinates, which parameterizes the coset $SL(4,\mathbb{R})/SO(4)$.
The map $f$ now identifies $SO(3,3)$ vectors $u$ with elements of the $SO(4)$ Lie algebra
according to
\begin{equation}
 f(u)=u_IT^I\; ,
\end{equation}
where $T^I$, with $I,J,\ldots = 1,\ldots ,6$ is a basis of
anti-symmetric $4\times 4$ matrices. We would like to choose these
matrices so that the first four, $T^1,\ldots ,T^4$ correspond to the
Pauli matrices $\sigma^1,\ldots , \sigma^4$ of the previous $N>2$
case, when written in real coordinates.  This ensures that our
result for $N=2$ indeed exactly reduces to the one for $N>2$ when
the additional degrees of freedom are ``switched off'' and,
hence, the action for both cases can be written in a uniform language.
It turns out that such a choice of matrices is given by
\begin{align}
T^{1}  &  =\left(
\begin{array}
[c]{cccc}%
0 & 0 & 0 & -1\\
0 & 0 & 1 & 0\\
0 & -1 & 0 & 0\\
1 & 0 & 0 & 0
\end{array}
\right)  ,\hspace{0.3cm}T^{2}=\left(
\begin{array}
[c]{cccc}%
0 & 0 & 1 & 0\\
0 & 0 & 0 & 1\\
-1 & 0 & 0 & 0\\
0 & -1 & 0 & 0
\end{array}
\right)  ,\hspace{0.3cm}\\
T^{3}  &  =\left(
\begin{array}
[c]{cccc}%
0 & -1 & 0 & 0\\
1 & 0 & 0 & 0\\
0 & 0 & 0 & 1\\
0 & 0 & -1 & 0
\end{array}
\right)  ,\hspace{0.3cm} T^{4}=\left(
\begin{array}
[c]{cccc}%
0 & -1 & 0 & 0\\
1 & 0 & 0 & 0\\
0 & 0 & 0 & -1\\
0 & 0 & 1 & 0
\end{array}
\right)\, .\hspace{0.3cm}%
\end{align}
The two remaining matrices can be taken as
\begin{equation}
T^{5}=\left(
\begin{array}
[c]{cccc}%
0 & 0 & 1 & 0\\
0 & 0 & 0 & -1\\
-1 & 0 & 0 & 0\\
0 & 1 & 0 & 0
\end{array}
\right)  ,\hspace{0.3cm}T^{6}=\left(
\begin{array}
[c]{cccc}%
0 & 0 & 0 & 1\\
0 & 0 & 1 & 0\\
0 & -1 & 0 & 0\\
-1 & 0 & 0 & 0
\end{array}
\right)  ,
\end{equation}
Note that $T^{1,2,3}$ and $T^{4,5,6}$ form the two sets of $SU(2)$ generators
within the $SO(4)$ Lie algebra. We may introduce a ``dual'' to the six $T^{I}$ matrices,
analogous to the definition of the $\bar{\sigma}^{I}$ matrices of the $N>2$ case, which will
prove useful in many calculations. We define
\begin{equation}
\left(  \overline{T}^{I}\right)  ^{AB}=-\frac{1}{2}\left(  T^{I}\right)
_{CD}\epsilon^{ABCD}%
\end{equation}
which has the simple form
\begin{equation}
(\overline{T}^{I})=(T^{u},-T^{\alpha}),
\end{equation}
where $u,v,\ldots =1,2,3$ and $\alpha ,\beta ,\ldots =4,5,6$. Indices $I,J,\ldots$ are raised
and lowered with the metric $(\eta_{IJ})={\rm diag}(-1,-1,-1,+1,+1,+1)$.
The matrices $T^{I}$ satisfy the following useful identities 
\begin{align}
\mathrm{tr}\left(  T^{I~}\overline{T}^{J}\right)   &  =4\eta^{IJ},\\
\left(  T^{I}\right)  _{AB}\left(  T^{J}\right)  _{CD}\eta_{IJ}  &
=2\epsilon_{ABCD},\\
\left(  T^{u}\right)  _{AB}\left(  T^{v}\right)  _{CD}\delta_{uv}  &
=\delta_{AC}\delta_{BD}-\delta_{AD}\delta_{BC}-\epsilon_{ABCD},\\
(T^{\alpha})_{AB}(T^{\beta})_{CD}\delta_{\alpha\beta}  &  =\delta_{AC}%
\delta_{BD}-\delta_{AD}\delta_{BC}+\epsilon_{ABCD}.
\end{align}
Key property of the map $f$ is
\begin{equation}
 (u_Iu^I)^2={\rm det}(f(u))
\end{equation}
for any $SO(3,3)$ vector $u_I$. This property can be used to show
that the map $F$ defined by
\begin{equation}
 F(v)u = f^{-1}\left( vf(u)v^T\right)
\end{equation}
is a group homeomorphism $F:SL(4,\mathbb{R})\rightarrow SO(3,3)$.
Solving for the $SO(3,3)$ images ${\ell_I}^{\ul{J}}={(F(v))_I}^{\ul{J}}$
one finds
\begin{equation}
 {\ell_I}^{\ul{J}}=\frac{1}{4}{\rm tr}\left( \bar{T}_IvT^Jv^T\right)\; .
\end{equation}
This induces the desired map between the cosets $SL(4,\mathbb{R})/SO(4)$
and $SO(3,3)/SO(3)^2$.


\section{Einstein-Yang-Mills supergravity in seven dimensions} \label{bigEYM}
In this section we will give a self-contained summary of minimal,
${\cal N}=1$ Einstein-Yang-Mills (EYM) supergravity in seven
dimensions. Although the theory may be formulated in two equivalent
ways, here we treat only the version in which the gravity multiplet
contains a three-form potential $C_{\mu\nu\rho}$ \cite{Park}, rather
than the dual formulation in terms of a two-index antisymmetric field
$B_{\mu\nu}$ which has been studied in
Refs.~\cite{Berghshoeff,Han}. This three-form formulation is better
suited for our application to M-theory. The theory has an $SU(2)$
rigid R-symmmetry that may be gauged and the resulting massive
theories were first constructed in
Refs.~\cite{Townsend,Mezincescu,Giani}. The seven-dimensional
supergravities we obtain by truncating M-theory are not massive and,
for this reason, we will not consider such theories with gauged
R-symmetry. The seven-dimensional pure supergravity theory can also be
coupled to $M$ vector multiplets~\cite{Park,Avramis,Cremmer,Nicolai},
transforming under a Lie group $G=U(1)^n\times H$, where $H$ is
semi-simple, in which case the vector multiplet scalars parameterize
the coset space $SO(3,M)/SO(3)\times SO(M)$. In this Appendix, we will
first review seven-dimensional ${\cal N}=1$ EYM supergravity with such
a gauge group $G$. This theory is used in the main part of the paper
to construct the complete action for low-energy M-theory on the
orbifolds $\mathbb{R}^{1,6}\times\mathbb{C}^{2}/\mathbb{Z}_{N}$. The
truncation of M-theory on these orbifolds to seven dimensions leads to
a $d=7$ EYM supergravity with gauge group $U(1)^n\times SU(N)$, where
$n=1$ for $N>2$ and $n=3$ for $N=2$. Here, the $U(1)^n$ part of the
gauge group originates from truncated bulk states, while the $SU(N)$
non-Abelian part corresponds to the additional states which arise on
the orbifold fixed plane.  Since we are constructing the coupled
11-/7-dimensional theory as an expansion in $SU(N)$ fields, the
crucial building block is a version of $d=7$ EYM supergravity with
gauge group $U(1)^n\times SU(N)$, expanded around the supergravity and
$U(1)^n$ part. This expanded version of the theory is presented in the
final part of this Appendix.

\subsection{General action and supersymmetry transformations} \label{full}
The field content of gauged $d=7$, ${\cal N}=1$ EYM supergravity
consists of two types of multiplets. The first, the gravitational
multiplet, contains a graviton $g_{\mu\nu}$ with associated vielbein ${e_\mu}^{\ul{\nu}}$, a gravitino $\psi_\mu^i$,
a symplectic Majorana spinor $\chi^i$, an $SU(2)$ triplet of Abelian
vector fields ${{A_\mu}^i}_j$ with field strengths ${F^i}_j=\mathrm{d}{A^i}_j$, a three form field $C_{\mu\nu\rho}$ with field strength $G=\mathrm{d}C$,
and a real scalar $\sigma$. So, in summary we have
\begin{equation} \label{gravmult}
 \left( g_{\mu\nu},C_{\mu\nu\rho},{{A_\mu}^i}_j,\sigma ,\psi^i_\mu ,\chi^i\right)\; .
\end{equation}
Here, $i,j,\ldots = 1,2$ are $SU(2)$ R-symmetry indices. The second type is the vector multiplet, which contains 
gauge vectors $A_\mu^a$ with field strengths $F^a={\cal D}A^a$, gauginos $\lambda^{ai}$ and $SU(2)$ triplets of real scalars ${\phi^{ai}}_j$. In summary, we have
\begin{equation} \label{vecmult}
 \left( A_\mu^a, {\phi^{ai}}_j, \lambda^{ai}\right)\; ,
\end{equation}
where $a,b,\ldots = 4,\ldots ,M+3$ are Lie algebra indices of the gauge group $G$.

It is sometimes useful to combine all vector fields, the three Abelian ones in the gravity
multiplet as well as the ones in the vector multiplets, into a single $SO(3,M)$
vector
\begin{equation}
 (A_\mu^{\tilde{I}})=\left({{A_\mu}^i}_j,A_\mu^a \right)\; ,
\end{equation}
where $\tilde{I},\tilde{J},\ldots = 1,\ldots ,M+3$. Under this combination, the corresponding field strengths are given by
\begin{equation}
F_{\mu\nu}^{\tilde{I}}=2\partial_{\lbrack\mu}A_{\nu]}^{\tilde{I}}+{f_{\tilde
{J}\tilde{K}}}^{\ti{I}}A_{\mu}^{\tilde{J}}A_{\nu}^{\tilde{K}},
\end{equation}
where ${f_{bc}}^a$ are the structure constants for $G$ and all other components of ${f_{\tilde{J}\tilde{K}}}^{\ti{I}}$ vanish. 

The coset space $SO(3,M)/SO(3)\times SO(M)$ is described by a
$(3+M)\times (3+M)$ matrix
$L_{\tilde{I}}^{\phantom{I}\underline{\tilde{J}}}$, which depends on
the $3M$ vector multiplet scalars and satisfies the $SO(3,M)$
orthogonality condition
\begin{equation}
L_{\tilde{I}}^{\phantom{I}  \underline{\tilde{J}}}L_{\tilde{K}}^{\phantom{K}
\underline{\tilde{L}}}\eta_{\underline{\tilde{J}}\underline{\tilde{L}}}%
=\eta_{\tilde{I}\tilde{K}}
\end{equation}
with
$(\eta_{\tilde{I}\tilde{J}})=(\eta_{\underline{\tilde{I}}\underline
{\tilde{J}}})=\mathrm{diag}(-1,-1,-1,+1,\ldots,+1)$. Here, indices
$\ti{I},\ti{J},\ldots = 1,\ldots,(M+3)$ transform under
$SO(3,M)$. Their flat counterparts $\ti{\ul{I}},\ti{\ul{J}},\ldots$
decompose into a triplet of $SU(2)$, corresponding to the
gravitational directions and $M$ remaining directions corresponding to
the vector multiplets. Thus we can write $L_{\tilde{I}}^{\phantom{I}
\underline{\tilde{J}}}\to\left(
{L_{\tilde{I}}}^u,L_{\tilde{I}}{}^{a}\right)$, where $u=1,2,3$. The
adjoint $SU(2)$ index $u$ can be converted into a pair of fundamental
$SU(2)$ indices by multiplication with the Pauli matrices, that is,
\begin{equation}
L_{\tilde{I}}{}^{i}{}_{j}=\frac{1}{\sqrt{2}}L_{\tilde{I}}{}^{u}\left(
\sigma_{u}\right) ^{i}{}_{j}\;.
\end{equation}
There are obviously many ways in
which one can parameterize the coset space $SO(3,M)/SO(3)\times SO(M)$
in terms of the physical vector multiplet scalar degrees of freedom
${{\phi_a}^i}_j$. A simple parameterization of this coset in terms
of $\Phi\equiv({\phi_{a}}^u)$ is given by
\begin{equation}
L_{\tilde{I}}^{\phantom{I}  \underline{\tilde{J}}}=\left(  \exp\left[
\begin{array}[c]{cc}%
0 & \Phi^T\\
\Phi & 0
\end{array}
\right]  \right)  _{\tilde{I}}^{\phantom{I}  \underline{\tilde{J}}}\text{.}%
\end{equation}
In the final paragraph of this appendix, when we expand seven-dimensional
supergravity, we will use a different parameterization, which is better 
adapted to this task. The Maurer-Cartan form of the matrix $L$, defined by
$L^{-1}\mathcal{D}L$, is needed to write down the theory. The
components $P$ and $Q$ are given explicitly by
\begin{align}
P_{\mu a\phantom{i} j}^{\phantom{\mu a} i} &  =
L_{\phantom{I} a}^{\tilde{I}}\left( \delta_{\tilde{I}}^{\tilde{K}}
 \partial_{\mu}+f_{\tilde{I}\tilde{J}}{}^{\tilde{K}}A_{\mu}^{\ti{J}}\right)
 L_{\tilde{K}\phantom{i} j}^{\phantom{K} i},\\
Q_{\mu\phantom{i} j}^{\phantom{\mu} i} &  =
L_{\phantom{Ii} k}^{\tilde{I}i}\left( \delta_{\tilde{I}}^{\tilde{K}}
 \partial_{\mu}+f_{\tilde{I}\tilde{J}}{}^{\tilde{K}}A_{\mu}^{\ti{J}}\right)
 L_{\tilde{K}\phantom{k} j}^{\phantom{K} k}.
\end{align}
The final ingredients needed are the following projections of the structure constants
\begin{align}
D  &  =if_{ab}^{\phantom{ab}  c}L_{\phantom{ai}  k}^{ai}L_{\phantom{bj}
i}^{bj}L_{c\phantom{k}  j}^{\phantom{c}  k},\nonumber\\
D_{\phantom{ai}  j}^{ai}  &  =if_{bc}^{\phantom{bc}  d}L_{\phantom{bi}
k}^{bi}L_{\phantom{ck}  j}^{ck}L_{d}^{\phantom{d}  a},\nonumber\\
D_{ab\phantom{i}  j}^{\phantom{ab}  i}  &  =f_{cd}^{\phantom{cd}  e}%
L_{a}^{\phantom{a}  c}L_{b}^{\phantom{b}  d}L_{e\phantom{i}  j}^{\phantom{e}
i}.
\end{align}

It is worth mentioning that invariance of the theory under the gauge
group $G$ and the R-symmetry group $SU(2)$ requires that the
Maurer-Cartan forms $P$ and $Q$ transform covariantly. It can be shown
that this is the case, if and only if the ``extended'' set of
structure constants $f_{\tilde{I}\tilde{J}}{}^{\tilde{K}}$ satisfy the
condition
\begin{equation}
f_{\tilde{I}\tilde{J}}{}^{\tilde{L}}\eta_{\tilde{L}}{}_{\tilde{K}%
}=f_{[\tilde{I}\tilde{J}}{}^{\tilde{L}}\eta{}_{\tilde{K}]}{}_{\tilde{L}%
}.\label{gauge condition}
\end{equation}
For any direct product factor of the total gauge group, this condition
can be satisfied in two ways. Either, the structure constants are
trivial, or the metric $\eta_{\tilde{I}\tilde{J}}$ is the Cartan-Killing
metric of this factor. In our particular case, the condition~\eqref{gauge condition}
is satisfied by making use of both these possibilities. 
The structure constants vanish for the ``gravitational'' part of the
gauge group and the $U(1)^n$ part within $G$. For the semi-simple part $H$
of $G$, one can always choose a basis, so its Cartan-Killing metric is
simply the Kronecker delta.

With everything in place, we now write down the Lagrangian for the
theory. Setting coupling constants to one, and neglecting four-fermi
terms, it is given by \cite{Park}
\begin{align}
e^{-1}\mathcal{L}_{\mathrm{YM}}  &  =\frac{1}{2}R-\frac{1}{2}\bar{\psi}_{\mu
}^{i}\Upsilon^{\mu\nu\rho}\hat{\mathcal{D}}_{\nu}\psi_{\rho i}-\frac{1}%
{96}e^{4\sigma}G_{\mu\nu\rho\sigma}G^{\mu\nu\rho\sigma}-\frac{1}{2}\bar{\chi
}^{i}\Upsilon^{\mu}\hat{\mathcal{D}}_{\mu}\chi_{i}-\frac{5}{2}\partial_{\mu
}\sigma\partial^{\mu}\sigma\nonumber\label{EYM}\\
&  +\frac{\sqrt{5}}{2}\left(  \bar{\chi}^{i}\Upsilon^{\mu\nu}\psi_{\mu i}%
+\bar{\chi}^{i}\psi_{i}^{\nu}\right)  \partial_{\nu}\sigma+e^{2\sigma}%
G_{\mu\nu\rho\sigma}\left[  \frac{1}{192}\left(  \bar{\psi}_{\lambda}%
^{i}\Upsilon^{\lambda\mu\nu\rho\sigma\tau}\psi_{\tau i}+12\bar{\psi}^{\mu
i}\Upsilon^{\nu\rho}\psi_{i}^{\sigma}\right)  \right. \nonumber\\
&  \hspace{4.2cm}\left.  +\frac{1}{48\sqrt{5}}\left(  4\bar{\chi}^{i}%
\Upsilon^{\mu\nu\rho}\psi_{i}^{\sigma}-\bar{\chi}^{i}\Upsilon^{\mu\nu
\rho\sigma\tau}\psi_{\tau i}\right)  -\frac{1}{320}\bar{\chi}^{i}\Upsilon
^{\mu\nu\rho\sigma}\chi_{i}\right] \nonumber\\
&  -\frac{1}{4}e^{-2\sigma}\left(  L_{\tilde{I}\phantom{i}  j}^{\phantom{I}
i}L_{\tilde{J}\phantom{j}  i}^{\phantom{J}  j}+L_{\tilde{I}}^{a}L_{\tilde{J}%
a}\right)  F_{\mu\nu}^{\tilde{I}}F^{\tilde{J}\mu\nu}-\frac{1}{2}\bar{\lambda
}^{ai}\Upsilon^{\mu}\hat{\mathcal{D}}_{\mu}\lambda_{ai}-\frac{1}{2}%
P_{\mu\phantom{ia}  j}^{\phantom{\mu}  ai}P_{\phantom{\mu}  a\phantom{j}
i}^{\mu\phantom{a}  j}\nonumber\\
&  -\frac{1}{\sqrt{2}}\left(  \bar{\lambda}^{ai}\Upsilon^{\mu\nu}\psi_{\mu
j}+\bar{\lambda}^{ai}\psi_{j}^{\nu}\right)  P_{\nu a\phantom{j}
i}^{\phantom{\nu a}  j}+\frac{1}{192}e^{2\sigma}G_{\mu\nu\rho\sigma}%
\bar{\lambda}^{ai}\Upsilon^{\mu\nu\rho\sigma}\lambda_{ai}\nonumber\\
&  -ie^{-\sigma}F_{\mu\nu}^{\tilde{I}}L_{\tilde{I}\phantom{j}  i}%
^{\phantom{I}  j}\left[  \frac{1}{4\sqrt{2}}\left(  \bar{\psi}_{\rho}%
^{i}\Upsilon^{\mu\nu\rho\sigma}\psi_{\sigma j}+2\bar{\psi}^{\mu i}\psi
_{j}^{\nu}\right)  +\frac{3}{20\sqrt{2}}\bar{\chi}^{i}\Upsilon^{\mu\nu}%
\chi_{j}\right. \nonumber\\
&  \hspace{4.2cm}\left.  -\frac{1}{4\sqrt{2}}\bar{\lambda}^{ai}\Upsilon
^{\mu\nu}\lambda_{aj}+\frac{1}{2\sqrt{10}}\left(  \bar{\chi}^{i}\Upsilon
^{\mu\nu\rho}\psi_{\rho j}-2\bar{\chi}^{i}\Upsilon^{\mu}\psi_{j}^{\nu}\right)
\right] \nonumber\\
&  +e^{-\sigma}F_{\mu\nu}^{\tilde{I}}L_{\tilde{I}a}\left[  \frac{1}{4}\left(
2\bar{\lambda}^{ai}\Upsilon^{\mu}\psi_{i}^{\nu}-\bar{\lambda}^{ai}%
\Upsilon^{\mu\nu\rho}\psi_{\rho i}\right)  +\frac{1}{2\sqrt{5}}\bar{\lambda
}^{ai}\Upsilon^{\mu\nu}\chi_{i}\right] \nonumber\\
&  +\frac{5}{180}e^{2\sigma}\left(  D^{2}-9D_{\phantom{ai}  j}^{ai}%
D_{a\phantom{j}  i}^{\phantom{a}  j}\right)  -\frac{i}{\sqrt{2}}e^{\sigma
}D_{ab\phantom{i}  j}^{\phantom{ab}  i}\bar{\lambda}^{aj}\lambda_{i}%
^{b}+\frac{i}{2}e^{\sigma}D_{a\phantom{i}  j}^{\phantom{a}  i}\left(
\bar{\psi}_{\mu}^{j}\Upsilon^{\mu}\lambda_{i}^{a}+\frac{2}{\sqrt{5}}\bar{\chi
}^{j}\lambda_{i}^{\phantom{i}  a}\right) \nonumber\\
&  +\frac{1}{60\sqrt{2}}e^{\sigma}D\left(  5\bar{\psi}_{\mu}^{i}\Upsilon
^{\mu\nu}\psi_{\nu i}+2\sqrt{5}\bar{\psi}_{\mu}^{i}\Upsilon^{\mu}\chi
_{i}+3\bar{\chi}^{i}\chi_{i}-5\bar{\lambda}^{ai}\lambda_{ai}\right)
\nonumber\\
&  -\frac{1}{96}\epsilon^{\mu\nu\rho\sigma\kappa\lambda\tau}C_{\mu\nu\rho
}F_{\sigma\kappa}^{\tilde{I}}F_{\tilde{I}}{}_{\lambda\tau}.
\end{align}
The covariant derivatives that appear here are given by 
\ba
\hat{\mathcal{D}}_\mu\psi_{\nu i}&=&\partial_{\mu}\psi_{\nu i}+\frac{1}{2}Q_{\mu
i}{}^{j}\psi_{\nu j}-\Gamma^\rho_{\mu\nu}\psi_{\rho i} +\frac{1}{4}\omega_{\mu}^{\phantom{\mu}  \underline
{\mu}\underline{\nu}}\Upsilon_{\underline{\mu}\underline{\nu}}\psi_{\nu i}, \\
\hat{\mathcal{D}}_{\mu}\chi_{i}&=&\partial_{\mu}\chi_{i}+\frac{1}{2}Q_{\mu
i}{}^{j}\chi_{j}+\frac{1}{4}\omega_{\mu}^{\phantom{\mu}  \underline
{\mu}\underline{\nu}}\Upsilon_{\underline{\mu}\underline{\nu}}\chi_{i},\\
\hat{\mathcal{D}}_{\mu}\la_{ai}&=&\partial_{\mu}\la_{ai}+\frac{1}{2}Q_{\mu
i}{}^{j}\la_{aj}+\frac{1}{4}\omega_{\mu}^{\phantom{\mu}  \underline
{\mu}\underline{\nu}}\Upsilon_{\underline{\mu}\underline{\nu}}\la_{ai}+{f_{ab}}^cA_\mu^b\la_{ci}.
\ea
The associated supersymmetry transformations, parameterized by the spinor $\varepsilon_i$, are, up to cubic fermion terms, given by
\begin{align}
\delta\sigma &  =\frac{1}{\sqrt{5}}\bar{\chi}^{i}\varepsilon_{i}\text{ ,
}\nonumber\\
{\delta e_{\mu}}^{\underline{\nu}}  &  =\bar{\varepsilon}^{i}\Upsilon
^{\underline{\nu}}\psi_{\mu i}\text{ ,}\nonumber
\end{align}
\begin{align}
\delta\psi_{\mu i}  &  =2\hat{\mathcal{D}}_{\mu}\varepsilon_{i}-\frac{1}%
{80}\left(  \Upsilon_{\mu}^{\phantom{\mu}  \nu\rho\sigma\eta}-\frac{8}%
{3}\delta_{\mu}^{\nu}\Upsilon^{\rho\sigma\eta}\right)  \varepsilon_{i}%
G_{\nu\rho\sigma\eta}e^{2\sigma}\text{ }\nonumber\\
&  +\frac{i}{5\sqrt{2}}\left(  \Upsilon_{\mu}^{\phantom{\mu}  \nu\rho}%
-8\delta_{\mu}^{\nu}\Upsilon^{\rho}\right)  \varepsilon_{j}F_{\nu\rho
}^{\tilde{I}}L_{\tilde{I}\phantom{i}  i}^{\phantom{I}  j}e^{-\sigma}%
-\frac{1}{15\sqrt{2}}e^{\sigma}\Upsilon_{\mu}\varepsilon_{i}D\text{
,}\nonumber\\
\delta\chi_{i}  &  =\sqrt{5}\Upsilon^{\mu}\varepsilon_{i}\partial_{\mu}%
\sigma-\frac{1}{24\sqrt{5}}\Upsilon^{\mu\upsilon\rho\sigma}\varepsilon
_{i}G_{\mu\nu\rho\sigma}e^{2\sigma}\text{ }-\frac{i}{\sqrt{10}}\Upsilon
^{\mu\nu}\varepsilon_{j}F_{\mu\nu}^{\tilde{I}}L_{\tilde{I}\phantom{i}
i}^{\phantom{I}  j}e^{-\sigma}+\frac{1}{3\sqrt{10}}e^{\sigma}\varepsilon
_{i}D\text{ ,}\nonumber\\
\delta C_{\mu\nu\rho}  &  =\left(  -3\bar{\psi}_{\left[  \mu\right.  }%
^{i}\Upsilon_{\left.  \nu\rho\right]  }\varepsilon_{i}-\frac{2}{\sqrt{5}}%
\bar{\chi}^{i}\Upsilon_{\mu\nu\rho}\varepsilon_{i}\right)  e^{-2\sigma}\text{
,} \label{7dsusy} \\
L_{\tilde{I}\phantom{i}  j}^{\phantom{I}  i}\delta A_{\mu}^{\tilde{I}}  &
=\left[  i\sqrt{2}\left(  \bar{\psi}_{\mu}^{i}\varepsilon_{j}-\frac{1}%
{2}\delta_{j}^{i}\bar{\psi}_{\mu}^{k}\varepsilon_{k}\right)  -\frac{2i}%
{\sqrt{10}}\left(  \bar{\chi}^{i}\Upsilon_{\mu}\varepsilon_{j}-\frac{1}{2}%
{}\delta_{j}^{i}\bar{\chi}^{k}\Upsilon_{\mu}\varepsilon_{k}\right)  \right]
e^{\sigma}\text{ ,}\nonumber\\
L_{\tilde{I}}^{\phantom{I}  a}\delta A_{\mu}^{\tilde{I}}  &  =\bar
{\varepsilon}^{i}\Upsilon_{\mu}\lambda_{i}^{a}e^{\sigma}\text{ ,}\nonumber\\
\delta L_{\tilde{I}\phantom{i}  j}^{\phantom{I}  i}  &  =-i\sqrt{2}%
\bar{\varepsilon}^{i}\lambda_{aj}L_{\tilde{I}}^{\phantom{I}  a}+\frac{i}%
{\sqrt{2}}\bar{\varepsilon}^{k}\lambda_{ak}L_{\tilde{I}}^{a}\delta_{j}%
^{i}\text{ ,}\nonumber\\
\delta L_{\tilde{I}}^{\phantom{I}  a}  &  =-i\sqrt{2}\bar{\varepsilon}%
^{i}\lambda_{j}^{a}L_{\tilde{I}\phantom{j}  i}^{\phantom{I}  j}\text{
,}\nonumber\\
\delta\lambda_{i}^{a}  &  =-\frac{1}{2}\Upsilon^{\mu\nu}\varepsilon_{i}%
F_{\mu\nu}^{\tilde{I}}L_{\tilde{I}}^{a}e^{-\sigma}+\sqrt{2}i\Upsilon^{\mu
}\varepsilon_{j}P_{\mu\phantom{aj}  i}^{\phantom{\mu }  aj}\text{ }-e^{\sigma
}\varepsilon_{j}D^{aj}{}_{i}\;. \nn 
\end{align}

\subsection{A perturbative expansion}\label{bigEYM2}
In this final section we expand the EYM supergravity of Section
\ref{full} around its supergravity and $U(1)^n$ part. The parameter
for the expansion is $h:=\kappa_7/g_\mathrm{YM}$, where $\kappa_7$ is
the coupling for gravity and $U(1)^n$ and $g_\mathrm{YM}$ is the
coupling for $H$, the non-Abelian part of the gauge group. To
determine the order in $h$ of each term in the Lagrangian, we need to
fix a convention for the energy dimensions of the fields. Within the
gravity and $U(1)$ vector multiplets, we assign energy dimension 0 to
bosonic fields and energy dimension 1/2 to fermionic fields.  For the
$H$ vector multiplet, we assign energy dimension 1 to the bosons and
3/2 to the fermions. With these conventions we can write
\begin{equation}
\mathcal{L}_{\mathrm{YM}}=\kappa_7^{-2}\left( \mathcal{L}_{(0)}+h^2\mathcal{L}_{(2)}+h^4\mathcal{L}_{(4)}+\ldots \right),
\end{equation}
where the $\mathcal{L}_{(m)}$, $m=0,2,4,\ldots$ are independent of
$h$. The first term in this series is the Lagrangian for EYM
supergravity with gauge group $U(1)^n$, whilst the second term
contains the leading order non-Abelian gauge multiplet terms. We will
write down these first two terms and provide truncated supersymmetry
transformation laws suitable for the theory at this order.

In order to carry out the expansion, it is necessary to cast the field
content in a form where the $H$ vector multiplet fields and the
gravity/$U(1)^n$ vector multiplet fields are disentangled. To this
end, we decompose the single Lie algebra indices
$a,b,\ldots=4,\ldots,M+3$ used in Section \ref{full} into indices
$\alpha,\beta,\ldots=4,\ldots,3+n$ that label the $U(1)$ directions
and redefined indices $a,b,\ldots=n+4,\ldots,M+3$ that are Lie algebra
indices of $H$. This makes the disentanglement straightforward for
most of the fields. For example, vector fields, which naturally
combine into the single entity $A_\mu^{\ti{I}}$, can simply be
decomposed as $A_\mu^{\ti{I}}=(A_\mu^I,A_\mu^a)$, where $A_\mu^I$,
$I=1,\ldots,n+3$, refers to the three vector fields in the gravity
multiplet and the $U(1)^n$ vector fields, and $A_\mu^a$ denotes the
$H$ vector fields. Similarly, the $U(1)$ gauginos are denoted by
$\la_{\alpha i}$, whilst the $H$ gauginos are denoted by
$\la_{ai}$. The situation is somewhat more complicated for the vector
multiplet scalar fields, which, as discussed, all together combine
into the single coset $SO(3,M)/SO(3)\times SO(M)$, parameterized by
the $SO(3,M)$ matrix $L$. It is necessary to find an explicit form for
$L$, which separates the $3n$ scalars in the $U(1)^n$ vector
multiplets from the $3(M-n)$ scalars in the $H$ vector multiplet. To
this end, we note that, in the absence of the $H$ states, the $U(1)^n$
states parameterize a $SO(3,n)/SO(3)\times SO(n)$ coset, described by
$(3+n)\times (3+n)$ matrices
${\ell_I}^{\underline{I}}=({\ell_I}^u,{\ell_I}^\alpha )$. Here,
$\ell\equiv (\ell_I^{\ph{I}u})$ are $(3+n)\times 3$ matrices where the
index $u=1,2,3$ corresponds to the three ``gravity'' directions and
$m\equiv ( \ell_I^{\ph{I}\alpha})$ are $(3+n)\times n$ matrices with
$\alpha =4,\ldots ,n+3$ labeling the $U(1)^n$ directions.  Let us
further denote the $SU(N)$ scalars by $\Phi\equiv
(\phi_a^{\ph{a}u})$. Then we can construct approximate representatives
$L$ of the large coset $SO(3,M)/SO(3)\times SO(M)$ by expanding, to
the appropriate order in $\Phi$, around the small coset
$SO(3,n)/SO(3)\times SO(n)$ represented by $\ell$ and $m$. Neglecting
terms of cubic and higher order in $\Phi$, this leads to
\begin{equation}\label{L}
L = \left( \begin{array}{ccc}
\ell+\frac{1}{2}h^2\ell\Phi^T\Phi & m & h\ell\Phi^T \\
h\Phi & 0 & \boldsymbol{1}_{M-n}+\frac{1}{2}h^2\Phi\Phi^T \\
\end{array} \right)\, .
\end{equation}
We note that the neglected $\Phi$ terms are of order $h^3$ and higher 
and, since we are aiming to construct the action only up to terms of order
$h^2$, are, therefore, not relevant in the present context. 

For the expansion of the action it is useful to re-write the coset parameterization~\eqref{L}
and the associated Maurer-Cartan forms $P$ and $Q$ in component form. We find
\begin{align}
{L_{I}{}^{i}}_{j}  &  ={\ell_{I}{}^{i}}_{j}+\frac{1}{2}h^2\ell_{I}{}^{k}{}_{l}\phi_{\phantom{bi}  k}^{al}\phi_{a}{}_{\phantom{bi}j}^{i}\; ,\\
{L_I}^\alpha &=h {\ell_I}^\alpha,  \\
L_{I}{}^{a}  &  =h\ell_{I}{}^{i}{}_{j}\phi_{\phantom{aj}i}^{aj},\\
L_{a}{}^{i}{}_{j}  &  =h\phi_{a}{}_{\phantom{b}j}^{i}\; ,\\
{L_a}^\alpha & = 0\; , \\
L_{a}{}^{b}  &  ={\delta_{a}}^{b}+\frac{1}{2}h^2\phi_{a}{}_{\phantom{i}j}^{i}
\phi^{bj}_{\phantom{bj}i}\; , \\
{P_{\mu\alpha}{}^{i}}_{j}  &  = {p_{\mu\alpha}{}^{i}}_{j}+\frac{1}{2}h^2{p_{\mu\alpha}}^{k}{}_{l}\phi^{a}{}^{l}{}_{k}{\phi_{a}{}^{i}}_{j}\; ,\\
{{P_{\mu a}}^i}_j  & = -h\hat{\mathcal{D}}_{\mu}{\phi_{a}{}^{i}}_{j}\; ,\\
Q_{\mu}{}^{i}{}_{j}  &  =q_{\mu}{}^{i}{}_{j}+\frac{1}{2}h^2\left( \phi^{a}{}^{i}{}%
_{k}\hat{\mathcal{D}}_{\mu}\phi_{a}{}^{k}{}_{j}-\phi_{a}{}^{k}{}_{j}\hat{\mathcal{D}}_{\mu}\phi^{a}{}^{i}{}_{k}\right)\; ,
\end{align}
where $p$ and $q$ are the Maurer-Cartan forms associated with the small coset matrix $\ell$. Thus
\begin{align}
p_{\mu\alpha\phantom{i}  j}^{\phantom{\mu\alpha }  i}  &  =\ell_{\phantom{I}
\alpha}^{I}\partial_{\mu}\ell_{I\phantom{i}  j}^{\phantom{\mu}  i}
,\\
q_{\mu\phantom{i}  j\phantom{k}  l}^{\phantom{\mu}  i\phantom{j}  k}  &
=\ell_{\phantom{Ii}  j}^{Ii}\partial_{\mu}\ell_{I\phantom{k}  l}%
^{\phantom{\mu}  k}\; ,\\
q_{\mu\phantom{i}  j}^{\phantom{\mu}  i}  &  =\ell_{\phantom{Ii}  k}%
^{Ii}\partial_{\mu}\ell_{I\phantom{k}  j}^{\phantom{\mu}  k}\; . %
\end{align}
The covariant derivative of the $H$ vector multiplet scalar ${{\phi_a}^i}_j$ is given by
\begin{equation}
\hat{\mathcal{D}}_{\mu}\phi_{a\phantom{i}j}^{\phantom{a}i}=\partial_{\mu
}\phi_{a\phantom{i}  j}^{\phantom{a}i}-q_{\mu\phantom{i}j\phantom{k}
l}^{\phantom{\mu}i\phantom{j}k}\phi_{a\phantom{l}  k}^{\phantom{a}
l}+f_{ab}^{\phantom{ab}c}A_{\mu}^{b}\phi_{c\phantom{i}j}^{\phantom{c}i}.
\end{equation}

Using the expressions above, it is straightforward to perform the expansion of $\mathcal{L}_{\mathrm{YM}}$ up to order $h^2\sim g_{\mathrm{YM}}^{-2}$. It is given by
\begin{align} \label{truncatedL}
\mathcal{L}_{\mathrm{YM}}\!  &  =\!\frac{1}{\kappa_{7}^{2}}\sqrt{-g}\left\{  \frac{1}{2}R-\frac{1}{2}\bar{\psi}_{\mu}^{i}\Upsilon^{\mu\nu\rho
}\hat{\mathcal{D}}_{\nu}\psi_{\rho i}-\frac{1}{4}e^{-2\sigma}\left(
\ell_{I\phantom{i}  j}^{\phantom{I}  i}\ell_{J\phantom{j}  i}^{\phantom{J}
j}+\ell_{I}^{\phantom{I}  \alpha}\ell_{J\alpha}\right)  F_{\mu\nu}^{I}%
F^{J\mu\nu}\right. \nonumber\\
&  \hspace{1.5cm}-\frac{1}{96}e^{4\sigma}G_{\mu\nu\rho\sigma}G^{\mu\nu\rho\sigma}-\frac{1}{2}\bar{\chi}^{i}\Upsilon^{\mu}\hat
{\mathcal{D}}_{\mu}\chi_{i}-\frac{5}{2}\partial_{\mu}\sigma\partial^{\mu
}\sigma+\frac{\sqrt{5}}{2}\left(  \bar{\chi}^{i}\Upsilon^{\mu\nu}\psi_{\mu
i}+\bar{\chi}^{i}\psi_{i}^{\nu}\right)  \partial_{\nu}\sigma\nonumber\\
&  \hspace{1.5cm}-\frac{1}{2}\bar{\lambda}^{\alpha i}\Upsilon^{\mu}%
\hat{\mathcal{D}}_{\mu}\lambda_{\alpha i}-\frac{1}{2}p_{\mu\alpha\phantom{i}
j}^{\phantom{\mu\alpha}  i}p_{\phantom{\mu\alpha j}  i}^{\mu\alpha j}%
-\frac{1}{\sqrt{2}}\left(  \bar{\lambda}^{\alpha i}\Upsilon^{\mu\nu}\psi_{\mu
j}+\bar{\lambda}^{\alpha i}\psi_{j}^{\nu}\right)  p_{\nu\alpha\phantom{j}
i}^{\phantom{\nu\alpha}  j}\nonumber\\
&  \hspace{1.5cm}+e^{2\sigma}G_{\mu\nu\rho\sigma}\left[  \frac{1}%
{192}\left(  12\bar{\psi}^{\mu i}\Upsilon^{\nu\rho}\psi_{i}^{\sigma}+\bar
{\psi}_{\lambda}^{i}\Upsilon^{\lambda\mu\nu\rho\sigma\tau}\psi_{\tau
i}\right)  +\frac{1}{48\sqrt{5}}\left(  4\bar{\chi}^{i}\Upsilon^{\mu\nu\rho
}\psi_{i}^{\sigma}\right.  \right. \nonumber\\
&  \hspace{3.5cm}\left.  \left.  -\bar{\chi}^{i}\Upsilon^{\mu\nu\rho\sigma
\tau}\psi_{\tau i}\right)  -\frac{1}{320}\bar{\chi}^{i}\Upsilon^{\mu\nu
\rho\sigma}\chi_{i}+\frac{1}{192}\bar{\lambda}^{\alpha i}\Upsilon^{\mu\nu
\rho\sigma}\lambda_{\alpha i}\right] \nonumber\\
&  \hspace{1.5cm}-ie^{-\sigma}F_{\mu\nu}^{I}\ell_{I\phantom{j}  i}%
^{\phantom{I}  j}\left[  \frac{1}{4\sqrt{2}}\left(  \bar{\psi}_{\rho}%
^{i}\Upsilon^{\mu\nu\rho\sigma}\psi_{\sigma j}+2\bar{\psi}^{\mu i}\psi
_{j}^{\nu}\right)  +\frac{1}{2\sqrt{10}}\left(  \bar{\chi}^{i}\Upsilon^{\mu
\nu\rho}\psi_{\rho j}-2\bar{\chi}^{i}\Upsilon^{\mu}\psi_{j}^{\nu}\right)
\right. \nonumber\\
&  \hspace{3.9cm}\left.  +\frac{3}{20\sqrt{2}}\bar{\chi}^{i}\Upsilon^{\mu\nu
}\chi_{j}-\frac{1}{4\sqrt{2}}\bar{\lambda}^{\alpha i}\Upsilon^{\mu\nu}%
\lambda_{\alpha j}\right] \nonumber\\
&  \hspace{1.5cm}+e^{-\sigma}F_{\mu\nu}^{I}\ell_{I\alpha}\left[  \frac{1}%
{4}\left(  2\bar{\lambda}^{\alpha i}\Upsilon^{\mu}\psi_{i}^{\nu}-\bar{\lambda
}^{\alpha i}\Upsilon^{\mu\nu\rho}\psi_{\rho i}\right)  +\frac{1}{2\sqrt{5}%
}\bar{\lambda}^{\alpha i}\Upsilon^{\mu\nu}\chi_{i}\right] \nonumber\\
&  \hspace{1.5cm}\left.  -\frac{1}{96}\epsilon^{\mu\nu\rho\sigma\kappa
\lambda\tau}C_{\mu\nu\rho}F_{\sigma\kappa}^{I}F_{I\lambda\tau}\right\} \nn \\
& +\frac{1}{g_{\mathrm{YM}}^{2}}\sqrt{-g}\left\{
-\frac{1}{4}e^{-2\sigma}F_{\mu\nu}^{a}F_{a}^{\mu\nu}-\frac{1}{2}%
\hat{\mathcal{D}}_{\mu}\phi_{a\phantom{i}  j}^{\phantom{a}  i}\hat
{\mathcal{D}}^{\mu}\phi_{\phantom{aj}  i}^{aj}-\frac{1}{2}\bar{\lambda}^{ai}\Upsilon^{\mu}\hat{\mathcal{D}}_{\mu}\lambda_{ai}\right. \nonumber\\
& \hspace{1.8cm} \left.  -e^{-2\sigma}\ell_{I\phantom{i}  j}^{\phantom{I}  i}%
\phi_{a\phantom{j}  i}^{\phantom{a}  j}F_{\mu\nu}^{I}F^{a\mu\nu}-\frac{1}%
{2}e^{-2\sigma}\ell_{I\phantom{i}  j}^{\phantom{I}  i}\phi_{a\phantom{j}
i}^{\phantom{a}  j}\ell_{J\phantom{k}  l}^{\phantom{J}  k}{\phi^{al}}_k
F_{\mu\nu}^{I}F_{\mu\nu}^{J}\right. \nonumber\\
& \hspace{1.8cm} -\frac{1}{2}p_{\mu\alpha\phantom{i}  j}^{\phantom{\mu\alpha}  i}%
\phi_{a\phantom{j}  i}^{\phantom{a}  j}p_{\phantom{\mu\alpha k}  l}^{\mu\alpha
k}\phi_{\phantom{al}  k}^{al}+\frac{1}{4}\phi_{a\phantom{i}  k}^{\phantom{a}
i}\hat{\mathcal{D}}_{\mu}{\phi^{ak}}_j\bar{\lambda
}^{\alpha j}\Upsilon^{\mu}\lambda_{\alpha i}\nonumber\\
& \hspace{1.8cm} -\frac{1}{\sqrt{2}}\left(  \bar{\lambda}^{\alpha i}\Upsilon^{\mu\nu}%
\psi_{\mu j}+\bar{\lambda}^{\alpha i}\psi_{j}^{\nu}\right)  \phi
_{a\phantom{j}  i}^{\phantom{a}  j}\phi_{\phantom{ak}  l}^{ak}p_{\nu
\alpha\phantom{l}  k}^{\phantom{\nu\alpha}  l}-\frac{1}{\sqrt{2}}\left(
\bar{\lambda}^{ai}\Upsilon^{\mu\nu}\psi_{\mu j}+\bar{\lambda}^{ai}\psi
_{j}^{\nu}\right)  \hat{\mathcal{D}}_{\nu}\phi_{a\phantom{j}  i}%
^{\phantom{ a}  j}\nonumber\\
& \hspace{1.8cm} +\frac{1}{192}e^{2\sigma}G_{\mu\nu\rho\sigma}\bar{\lambda}%
^{ai}\Upsilon^{\mu\nu\rho\sigma}\lambda_{ai}+\frac{i}{4\sqrt{2}}e^{-\sigma
}F_{\mu\nu}^{I}\ell_{I\phantom{j}  i}^{\phantom{I}  j}\bar{\lambda}%
^{ai}\Upsilon^{\mu\nu}\lambda_{aj}\nonumber\\
& \hspace{1.8cm} -\frac{i}{2}e^{-\sigma}\left(  F_{\mu\nu}^{I}\ell_{I\phantom{k}
l}^{\phantom{I}  k}\phi_{\phantom{al}  k}^{al}\phi_{a\phantom{i}
j}^{\phantom{a}  i}+2F_{\mu\nu}^{a}\phi_{a\phantom{j}  i}^{\phantom{a}
j}\right)  \left[  \frac{1}{4\sqrt{2}}\left(  \bar{\psi}_{\rho}^{i}%
\Upsilon^{\mu\nu\rho\sigma}\psi_{\sigma j}+2\bar{\psi}^{\mu i}\psi_{j}^{\nu
}\right)  \right. \nonumber\\
& \hspace{1.8cm} \left.  +\frac{3}{20\sqrt{2}}\bar{\chi}^{i}\Upsilon^{\mu\nu}\chi
_{j}-\frac{1}{4\sqrt{2}}\bar{\lambda}^{\alpha i}\Upsilon^{\mu\nu}%
\lambda_{\alpha j}+\frac{1}{2\sqrt{10}}\left(  \bar{\chi}^{i}\Upsilon^{\mu
\nu\rho}\psi_{\rho j}-2\bar{\chi}^{i}\Upsilon^{\mu}\psi_{j}^{\nu}\right)
\right] \nonumber\\
& \hspace{1.8cm} +e^{-\sigma}F_{\mu\nu}^{a}\left[  \frac{1}{4}\left(  2\bar{\lambda}%
^{ai}\Upsilon^{\mu}\psi_{i}^{\nu}-\bar{\lambda}^{ai}\Upsilon^{\mu\nu\rho}%
\psi_{\rho i}\right)  +\frac{1}{2\sqrt{5}}\bar{\lambda}^{ai}\Upsilon^{\mu\nu
}\chi_{i}\right] \nonumber\\
& \hspace{1.8cm} +\frac{1}{4}e^{2\sigma}f_{bc}^{\phantom{bc}  a}f_{dea}\phi_{\phantom{bi}
k}^{bi}\phi_{\phantom{ck}  j}^{ck}\phi_{\phantom{dj}  l}^{dj}\phi
_{\phantom{el}  i}^{el}-\frac{1}{2}e^{\sigma}f_{abc}\phi_{\phantom{bi}
k}^{bi}\phi_{\phantom{ck}  j}^{ck}\left(  \bar{\psi}_{\mu}^{j}\Upsilon^{\mu
}\lambda_{i}^{a}+\frac{2}{\sqrt{5}}\bar{\chi}^{j}\lambda_{i}^{\phantom{i}
a}\right) \nonumber\\
& \hspace{1.8cm} -\frac{i}{\sqrt{2}}e^{\sigma}f_{ab}^{\phantom{ab}  c}\phi_{c\phantom{i}
j}^{\phantom{c}  i}\bar{\lambda}^{aj}\lambda_{i}^{b}+\frac{i}{60\sqrt{2}%
}e^{\sigma}f_{ab}^{\phantom{ab}  c}\phi_{\phantom{al}  k}^{al}\phi
_{\phantom{bj}  l}^{bj}\phi_{c\phantom{k}  j}^{\phantom{c}  k}\left(
5\bar{\psi}_{\mu}^{i}\Upsilon^{\mu\nu}\psi_{\nu i}+2\sqrt{5}\bar{\psi}_{\mu
}^{i}\Upsilon^{\mu}\chi_{i}\right. \nonumber
\end{align}
\begin{align}
& \hspace{1.8cm} \left.  \left.  +3\bar{\chi}^{i}\chi_{i}-5\bar{\lambda}^{\alpha i}%
\lambda_{\alpha i}\right)  -\frac{1}{96}\epsilon^{\mu\nu\rho\sigma
\kappa\lambda\tau}C_{\mu\nu\rho}F_{\sigma\kappa}^{a}F_{a\lambda\tau}\right\}.
\end{align}
The associated supersymmetry transformations have an expansion similar
to that of the Lagrangian. Thus, the supersymmetry transformation of a
field $X$ takes the form
\begin{equation}
\delta X = \delta^{(0)}X+h^2\delta^{(2)}X+h^4\delta^{(4)}X+\ldots\;.
\end{equation}
We give the first two terms of this series for the gravity and $U(1)$
vector multiplet fields, and just the first term for the $H$ vector
multiplet fields. These terms are precisely those required to prove
that the Lagrangian given in Eq. \eqref{truncatedL} is supersymmetric
to order $h^2\sim g^{-2}_{\mathrm{YM}}$. They are
\begin{align}
\delta\sigma &  =\frac{1}{\sqrt{5}}\bar{\chi}^{i}\varepsilon_{i},\nonumber\\
\delta {e_{\mu}}^{\underline{\nu}}  &  =\bar{\varepsilon}^{i}%
\Upsilon^{\underline{\nu}}\psi_{\mu i},\nonumber\\
\delta\psi_{\mu i}  &  =2\hat{\mathcal{D}}_{\mu}\varepsilon_{i}-\frac{1}%
{80}\left(  \Upsilon_{\mu}^{\phantom{\mu}  \nu\rho\sigma\eta}-\frac{8}%
{3}\delta_{\mu}^{\nu}\Upsilon^{\rho\sigma\eta}\right)  \varepsilon_{i}%
G_{\nu\rho\sigma\eta}e^{2\sigma}+\frac{i}{5\sqrt{2}}\left(
\Upsilon_{\mu}^{\phantom{\mu}  \nu\rho}-8\delta_{\mu}^{\nu}\Upsilon^{\rho
}\right)  \varepsilon_{j}F_{\nu\rho}^{I}\ell_{I\phantom{i}  i}^{\phantom{I}
j}e^{-\sigma}\nonumber\\
&  +\frac{\kappa_{7}^{2}}{g_{\mathrm{YM}}^{2}}\left\{  \frac{1}{2}\left(
\phi_{ak}^{\phantom{ak}  j}\hat{\mathcal{D}}_{\mu}\phi_{\phantom{a}
i}^{a\phantom{i}  k}-\phi_{\phantom{a}  i}^{a\phantom{i}  k}\hat{\mathcal{D}%
}_{\mu}\phi_{ak}^{\phantom{ak}  j}\right)  \varepsilon_{j}-\frac{i}{15\sqrt
{2}}\Upsilon_{\mu}\varepsilon_{i}f_{ab}^{\phantom{ab}  c}\phi_{\phantom{al}
k}^{al}\phi_{\phantom{bj}  l}^{bj}\phi_{c\phantom{k}  j}^{\phantom{c}
k}e^{\sigma}\right. \nonumber\\
&  \left.  \hspace{1.1cm}+\frac{i}{10\sqrt{2}}\left(  \Upsilon_{\mu
}^{\phantom{\mu}  \nu\rho}-8\delta_{\mu}^{\nu}\Upsilon^{\rho}\right)
\varepsilon_{j}\left(  F_{\nu\rho}^{I}\ell_{I\phantom{k}  l}^{\phantom{I}
k}\phi_{\phantom{al}  k}^{al}\phi_{a\phantom{j}  i}^{\phantom{a}  j}%
+2F_{\nu\rho}^{a}\phi_{a\phantom{j}  i}^{\phantom{a}  j}\right)  e^{-\sigma
}\right\}  ,\nonumber\\
\delta\chi_{i}  &  =\sqrt{5}\Upsilon^{\mu}\varepsilon_{i}\partial_{\mu}%
\sigma-\frac{1}{24\sqrt{5}}\Upsilon^{\mu\upsilon\rho\sigma}\varepsilon
_{i}G_{\mu\nu\rho\sigma}e^{2\sigma}\text{ }-\frac{i}{\sqrt{10}%
}\Upsilon^{\mu\nu}\varepsilon_{j}F_{\mu\nu}^{I}\ell_{I\phantom{i}
i}^{\phantom{I}  j}e^{-\sigma}\nonumber\\
&  +\frac{\kappa_{7}^{2}}{g_{\mathrm{YM}}^{2}}\left\{  -\frac{i}{2\sqrt{10}%
}\Upsilon^{\mu\nu}\varepsilon_{j}\left(  F_{\mu\nu}^{I}\ell_{I\phantom{k}
l}^{\phantom{I}  k}\phi_{\phantom{al}  k}^{al}\phi_{a\phantom{j}
i}^{\phantom{a}  j}+2F_{\mu\nu}^{a}\phi_{a\phantom{j}  i}^{\phantom{a}
j}\right)  e^{-\sigma}\right. \nonumber\\
&  \left.  +\frac{i}{3\sqrt{10}}\varepsilon_{i}f_{ab}^{\phantom{ab}  c}%
\phi_{\phantom{al}  k}^{al}\phi_{\phantom{bj}  l}^{bj}\phi_{c\phantom{k}
j}^{\phantom{c}  k}e^{\sigma}\right\}  ,\nonumber\\
\delta C_{\mu\nu\rho}  &  =\left(  -3\bar{\psi}_{\left[  \mu\right.
}^{i}\Upsilon_{\left.  \nu\rho\right]  }\varepsilon_{i}-\frac{2}{\sqrt{5}}%
\bar{\chi}^{i}\Upsilon_{\mu\nu\rho}\varepsilon_{i}\right)  e^{-2\sigma
},\nonumber\\
\ell_{I\phantom{i}  j}^{\phantom{I}  i}\delta A_{\mu}^{I}  &  =\left[
i\sqrt{2}\left(  \bar{\psi}_{\mu}^{i}\varepsilon_{j}-\frac{1}{2}\delta_{j}%
^{i}\bar{\psi}_{\mu}^{k}\varepsilon_{k}\right)  -\frac{2i}{\sqrt{10}}\left(
\bar{\chi}^{i}\Upsilon_{\mu}\varepsilon_{j}-\frac{1}{2}{}\delta_{j}^{i}%
\bar{\chi}^{k}\Upsilon_{\mu}\varepsilon_{k}\right)  \right]  e^{\sigma
}\nonumber\\
&  +\frac{\kappa_{7}^{2}}{g_{\mathrm{YM}}^{2}}\left\{  \left(  \frac{i}%
{\sqrt{2}}\bar{\psi}_{\mu}^{k}\varepsilon_{l}-\frac{i}{\sqrt{10}}\bar{\chi
}^{k}\Upsilon_{\mu}\varepsilon_{l}\right)  \phi_{\phantom{al}  k}^{al}%
\phi_{a\phantom{i}  j}^{\phantom{a}  i}e^{\sigma}-\bar{\varepsilon}%
^{k}\Upsilon_{\mu}\lambda_{k}^{a}\phi_{a\phantom{i}  j}^{\phantom{a}
i}e^{\sigma}\right\}  ,\\
\ell_{I}^{\phantom{I}  \alpha}\delta A_{\mu}^{I}  &  =\bar{\varepsilon}%
^{i}\Upsilon_{\mu}\lambda_{i}^{\alpha}e^{\sigma},\nonumber\\
\delta\ell_{I\phantom{i}  j}^{\phantom{I}  i}  &  =-i\sqrt{2}\bar{\varepsilon
}^{i}\lambda_{\alpha j}\ell_{I}^{\phantom{I}  \alpha}+\frac{i}{\sqrt{2}}%
\bar{\varepsilon}^{k}\lambda_{\alpha k}\ell_{I}^{\phantom{I}  \alpha}%
\delta_{j}^{i}\nonumber\\
&  +\frac{\kappa_{7}^{2}}{g_{\mathrm{YM}}^{2}}\left\{  \frac{i}{\sqrt{2}%
}\left[  \bar{\varepsilon}^{k}\lambda_{\alpha l}\phi_{\phantom{al}  k}%
^{al}\phi_{a\phantom{i}  j}^{\phantom{a}  i}\ell_{I}^{\phantom{I}  \alpha
}+\bar{\varepsilon}^{l}\lambda_{ak}\phi_{\phantom{ai}  j}^{ai}\ell
_{I\phantom{k}  l}^{\phantom{I}  k}-\left(  \bar{\varepsilon}^{i}\lambda
_{aj}-\frac{1}{2}\delta_{j}^{i}\bar{\varepsilon}^{m}\lambda_{am}\right)
\phi_{\phantom{al}  k}^{al}\ell_{I\phantom{k}  l}^{\phantom{I}  k}\right]
\right\}  ,\nonumber\\
\delta\ell_{I}^{\phantom{I}  \alpha}  &  =-i\sqrt{2}\bar{\varepsilon}%
^{i}\lambda_{j}^{\alpha}\ell_{I\phantom{j}  i}^{\phantom{I}  j}+\frac{\kappa
_{7}^{2}}{g_{\mathrm{YM}}^{2}}\left\{  -\frac{i}{\sqrt{2}}\bar{\varepsilon
}^{i}\lambda_{j}^{\alpha}\phi_{\phantom{aj}  i}^{aj}\phi_{a\phantom{l}
k}^{\phantom{a}  l}\ell_{I\phantom{k}  l}^{\phantom{I}  k}\right\}
,\nonumber\\
\delta\lambda_{i}^{\alpha}  &  =-\frac{1}{2}\Upsilon^{\mu\nu}\varepsilon
_{i}F_{\mu\nu}^{I}\ell_{I}^{\phantom{I}  \alpha}e^{-\sigma}+\sqrt{2}%
i\Upsilon^{\mu}\varepsilon_{j}p_{\mu\phantom{\alpha j}  i}^{\phantom{\mu}
\alpha j}+\frac{\kappa_{7}^{2}}{g_{\mathrm{YM}}^{2}}\left\{  \frac{i}{\sqrt
{2}}\Upsilon^{\mu}\varepsilon_{j}\phi_{ai}^{\phantom{ai}  j}p_{\mu
\phantom{\alpha k}  l}^{\phantom{\mu}  \alpha k}\phi_{\phantom{al}  k}%
^{al}\right\}  ,\nonumber
\end{align}
\begin{align}
\delta A_{\mu}^{a}  &  =\bar{\varepsilon}^{i}\Upsilon_{\mu}\lambda_{i}%
^{a}e^{\sigma}-\left(  i\sqrt{2}\psi_{\mu}^{i}\varepsilon_{j}-\frac{2i}%
{\sqrt{10}}\bar{\chi}^{i}\Upsilon_{\mu}\varepsilon_{j}\right)  \phi
_{\phantom{aj}  i}^{aj}e^{\sigma},\nonumber\\
\delta\phi_{a\phantom{i}  j}^{\phantom{a}  i}  &  =-i\sqrt{2}\left(
\bar{\varepsilon}^{i}\lambda_{aj}-\frac{1}{2}\delta_{j}^{i}\bar{\varepsilon
}^{k}\lambda_{ak}\right)  ,\nonumber\\
\delta\lambda_{i}^{a}  &  =\left(  -\frac{1}{2}\Upsilon^{\mu\nu}%
\varepsilon_{i}\left(  F_{\mu\nu}^{I}\ell_{I\phantom{j}  k}^{\phantom{I}
j}\phi_{\phantom{ak}  j}^{ak}+F_{\mu\nu}^{a}\right)  e^{-\sigma}-i\sqrt
{2}\Upsilon^{\mu}\varepsilon_{j}\hat{\mathcal{D}}_{\mu}\phi_{\phantom{a}
i}^{a\phantom{i}  j}-i\varepsilon_{j}f_{\phantom{a}  bc}^{a}\phi
_{\phantom{bj}  k}^{bj}\phi_{\phantom{ck}  i}^{ck}\right), \nn\\
\delta A_{\mu}^{a}  &  =\bar{\varepsilon}^{i}\Upsilon_{\mu}\lambda_{i}%
^{a}e^{\sigma}-\left(  i\sqrt{2}\psi_{\mu}^{i}\varepsilon_{j}-\frac{2i}%
{\sqrt{10}}\bar{\chi}^{i}\Upsilon_{\mu}\varepsilon_{j}\right)  \phi
_{\phantom{aj}  i}^{aj}e^{\sigma},\nonumber\\
\delta\phi_{a\phantom{i}  j}^{\phantom{a}  i}  &  =-i\sqrt{2}\left(
\bar{\varepsilon}^{i}\lambda_{aj}-\frac{1}{2}\delta_{j}^{i}\bar{\varepsilon
}^{k}\lambda_{ak}\right)  ,\nonumber\\
\delta\lambda_{i}^{a}  &  =-\frac{1}{2}\Upsilon^{\mu\nu}\varepsilon_{i}\left(
F_{\mu\nu}^{I}\ell_{I\phantom{j}  k}^{\phantom{I}  j}\phi_{\phantom{ak}
j}^{ak}+F_{\mu\nu}^{a}\right)  e^{-\sigma}-i\sqrt{2}\Upsilon^{\mu}%
\varepsilon_{j}\hat{\mathcal{D}}_{\mu}\phi_{\phantom{a}  i}^{a\phantom{i}
j}-i\varepsilon_{j}f_{\phantom{a}  bc}^{a}\phi_{\phantom{bj}  k}^{bj}%
\phi_{\phantom{ck}  i}^{ck}.\nonumber
\end{align}
This completes our review of $\mathcal{N}=1$ EYM supergravity in seven dimensions.


\end{document}